\documentclass[a4paper,11pt]{article}
\usepackage{jheppub} 
\usepackage{lineno}
\usepackage{graphicx,amsfonts,amsmath,amssymb,epsfig,color, mathrsfs,latexsym,url,wasysym,float,xfrac}

\usepackage{amsmath,amssymb}
\usepackage{tikz}
\usetikzlibrary{arrows.meta,calc}
\DeclareMathOperator{\Tr}{Tr}
\DeclareMathOperator{\End}{End}
\DeclareMathOperator{\Hom}{Hom}

\usepackage{etoolbox}
\makeatletter
\patchcmd{\l@section}
  {\addvspace{1.0em \@plus\p@}}
  {\addvspace{0.34em \@plus\p@}}
  {}
  {\PackageWarning{toc-spacing}{Section spacing patch failed}}
\makeatother
\renewcommand{\afterTocRuleSpace}{\clearpage}

\newcommand{\be}{\begin{equation}}
\newcommand{\ee}{\end{equation}}

\newcommand{\ba}{\begin{eqnarray}}
\newcommand{\ea}{\end{eqnarray}}

\newcommand{\beq}{\begin{equation}}  \newcommand{\eeq}{\end{equation}}

\newcommand{\M}{\mathcal{M}}

\newcommand{\g}{\mathfrak{g}}

\newcommand{\nah}{\text{NAH}}
\newcommand{\geom}{\text{geom}}

\numberwithin{equation}{section}

\definecolor{hepblue}{RGB}{0,40,145}
\definecolor{branchred}{RGB}{170,20,20}
\definecolor{lightublue}{RGB}{232,241,250}
\definecolor{lightvred}{RGB}{250,235,235}

\tikzset{
  axis/.style={-{Latex[length=2.2mm,width=1.5mm]}, line width=0.6pt},
  ubranch/.style={hepblue, line width=1pt},
  vbranch/.style={branchred, line width=1pt},
  dblue/.style={hepblue, dashed, line width=0.7pt},
  dred/.style={branchred, dashed, line width=0.7pt},
  bluearrow/.style={-{Latex[length=3mm,width=2mm]}, line width=1pt, hepblue},
  redarrow/.style={-{Latex[length=3mm,width=2mm]}, line width=1pt, branchred},
  mainarrow/.style={-{Latex[length=4mm,width=2.5mm]}, line width=1.2pt}
}

\title{Asymptotic limits in class-\texorpdfstring{$\mathcal S$}{S} theories and non-Abelian Hodge theory}

\author{Thomas W.~Grimm\ $^{a,b}$,}
\author{Amineh Mohseni\ $^c$}

\affiliation[a]{Institute for Theoretical Physics, Utrecht University,\\
Princetonplein 5, 3584 CC Utrecht, 
The Netherlands}

\affiliation[b]{Center of Mathematical Sciences and Applications,\\
Harvard University, Cambridge, MA 02138, USA }

\affiliation[c]{Jefferson Physical Laboratory, Harvard University,\\ 
Cambridge, MA 02138, USA}

\emailAdd{t.w.grimm@uu.nl, amohseni@g.harvard.edu}

\abstract{
We develop a non-Abelian Hodge-theoretic refinement of asymptotic
limits in type-$A$ class-$\mathcal S$ theories, enriching the
geometric description of the degenerating UV curve with
gauge-theoretic data encoded by the associated Hitchin system.
A degeneration of the UV curve
produces a long plumbing tube, which we equip with the local monodromy data of
a Hitchin--Simpson flat connection admitting a regular-singular logarithmic
model. The semisimple and unipotent parts of the monodromy govern,
respectively, the power-law and logarithmic growth of flat sections
through the tube. Combining this non-Abelian holonomy with the geometric
Picard--Lefschetz monodromy yields a decorated cusp label that incorporates
Higgs-bundle information and extends tube-wise to intersections of
boundary divisors.
We also discuss a particular class of monodromy labels for which
the weak gauge algebra specified by the fixture and gluing data
lies in the reductive monodromy centralizer.
In type $A$, we organize the local monodromy labels into discrete types
specified by the eigenspace multiplicities of the semisimple part and
the Jordan type of the nilpotent logarithm of the unipotent part.
Finally, we illustrate the construction for $SL(3,\mathbb C)$
on the four-punctured sphere and $SL(4,\mathbb C)$
on the two-punctured torus.
}

\begin{document}
\maketitle

\section{Introduction}

The low-energy dynamics of a supersymmetric quantum field theory depends not only on its couplings, but also on the choice of vacuum. Consequently, approaching a boundary point in coupling space does not by itself determine the asymptotic low-energy physics: different choices of vacuum data may exhibit distinct behavior along the same degeneration. This suggests that weak-coupling limits should be characterized by more than the limiting behavior of the couplings alone. In class-$\mathcal S$ theories \cite{Gaiotto:2009we}, the complex structure of the UV curve parametrizes the exactly marginal couplings, while the associated Hitchin system encodes the Coulomb-branch geometry and vacuum data. We show that \textit{non-Abelian Hodge theory} provides a natural framework for relating these structures. Through the non-Abelian Hodge correspondence, a chosen Higgs bundle determines a flat connection whose local monodromy can be followed as the UV curve degenerates. In this way, the description of a conformal-manifold cusp can be refined by geometric and non-Abelian monodromy data that retain information about the accompanying vacuum.

The degeneration of the UV curve produces a long plumbing tube carrying
the chosen Higgs-bundle data and their associated
\emph{Hitchin--Simpson flat connection}.
As a tube pinches, its two ends can be viewed as additional punctures.
Near these ends, we take the flat connection to have logarithmic
singularities in the UV-curve coordinates.
We therefore consider families admitting the regular-singular tube model described below, with the 
flat-connection data specified separately from the complex-structure
degeneration. Within this setting, we study the relevant monodromies,
together with the transport and asymptotic growth of flat sections
through the long tube.

The relation between class-\(\mathcal S\) theories, Hitchin systems, and
moduli spaces of flat connections has appeared in several complementary
forms. In particular, the spectral-network constructions of
Gaiotto, Moore, and Neitzke relate non-Abelian flat connections on the UV
curve to Abelian flat connections on the associated spectral cover and
provide natural coordinates on moduli spaces of flat connections
\cite{GaiottoMooreNeitzkeSpectralNetworks,
GaiottoMooreNeitzkeSpectralNetworksSnakes}.
From a complementary global perspective, Balasubramanian, Distler, and
Donagi studied families of Hitchin systems with prescribed puncture data as the UV curve
varies over the Deligne--Mumford moduli space, including their degeneration
to nodal curves \cite{Balasubramanian:2020fwc}. In F-theory compactifications, limiting mixed Hodge structures
and local Hitchin systems have also been used to study T-brane backgrounds~\cite{AndersonHeckmanKatz2013, AndersonHeckmanKatzSchaposnik2017}.
Our use of these structures is different: we employ the non-Abelian Hodge correspondence to associate local de Rham monodromy data of a
Hitchin--Simpson flat connection directly with the plumbing tubes appearing at conformal-manifold boundary divisors, and combine these data with the geometric monodromy of the degenerating UV curve, to propose a refined characterization of the cusp.

In ordinary Hodge theory, Deligne's mixed Hodge structures extend
the cohomological description to singular and noncompact algebraic
varieties \cite{Deligne1971HodgeII,Deligne1974HodgeIII}.
For degenerating families, 
the asymptotic behavior of the Hodge structure and its associated period data is encoded by 
monodromy and limiting mixed Hodge
structures \cite{Schmid1973,Kaplan1986}. 
These methods have also been used to study infinite-distance limits and light
towers in Calabi--Yau compactifications
\cite{Grimm:2018ohb,Grimm:2018cpv,Corvilain:2018lgw, Monnee:2025ComplexStructureLimits,Hassfeld:2025EmergentStrings}
and construct general asymptotic periods
\cite{Bastian:2021eom}.
Non-Abelian Hodge theory, on the other hand, relates Higgs bundles, flat connections,
and representations of the fundamental group through harmonic
metrics, under appropriate stability and topological conditions
\cite{Donaldson1987,Hitchin1987SelfDuality,Corlette1988flat,
Simpson1992HiggsBA}.
On punctured curves, this correspondence extends to the tame
harmonic-bundle setting, with filtered or parabolic data encoding
the behavior at the punctures
\cite{simpson1990harmonic,mochizuki2004kobayashi,
mochizuki2009kobayashi}.
The associated local asymptotic theory relates monodromy and
filtration data to the growth of sections
\cite{Mochizuki2002,Mochizuki2003Asymptotic}, providing the
analytic framework for our study of regular-singular
flat connections on degenerating plumbing tubes.

Non-Abelian Hodge theory therefore provides a natural approach
to characterizing cusps in class-\(\mathcal S\) theories.
We augment the geometric monodromy data of the degenerating UV curve
with the local monodromy data of a Hitchin--Simpson flat connection. Compared
with the ordinary Hodge-theoretic description of Calabi--Yau
degenerations, this introduces an \emph{additional non-Abelian
structure} carried by the Higgs bundle and its associated flat
connection. In type \(A_{N-1}\), the possible local monodromy types
are organized into discrete types by semisimple eigenspace multiplicities and nilpotent
Jordan partitions within the \(N\)-dimensional defining
representation. The resulting set of local monodromy decoration types grows
with \(N\).

We first formulate the regular-singular tube model for a one-parameter
degeneration and extend it tube by tube to intersections of boundary
divisors, deriving the associated transfer matrices and their growth. 
A loop around a boundary divisor induces a Dehn twist on the UV curve.
We associate to this loop the local non-Abelian holonomy around the
corresponding tube core, together with the Picard--Lefschetz monodromy
of the curve. We then specialize
the construction to type-\(A\) class-\(\mathcal S\) conformal manifolds. We illustrate the resulting single- and multi-divisor labels
using two examples: \(SL(3,\mathbb C)\) on the four-punctured sphere
and \(SL(4,\mathbb C)\) on the two-punctured torus.
We then discuss a particular class of monodromy labels, selected
by imposing that the weak gauge algebra specified independently
by the fixture and gluing data lie in the corresponding compact
reductive monodromy centralizer.

The non-Abelian Hodge-theoretic refinement of weak-coupling cusps developed here suggests two natural directions for further investigation. The first concerns the CFT Distance Conjecture, which relates infinite-distance limits of conformal manifolds to the emergence of higher-spin symmetry \cite{Baume:2020dqd,Perlmutter:2020buo}, together with its subsequent developments \cite{Baume:2023msm,Calderon-Infante:2024oed,CalderonInfante:2026quivers,MantegazzaEtAl2026,
BaumeMantegazza2026,Fenati:2026hkn}. In class-$\mathcal S$ theories, our construction provides additional monodromy data associated with a given weak-coupling degeneration, and it is natural to ask whether this refinement can help distinguish or organize the asymptotic regimes relevant to the Distance Conjecture. A second direction concerns a possible connection with Geometric Langlands. The moduli spaces of Higgs bundles and flat connections that enter our construction also play a central role in gauge-theoretic approaches to Geometric Langlands \cite{Kapustin:2006pk,Gukov:2006jk,Witten:2009at}. This raises the question of whether the local monodromy data and asymptotic structures identified here admit a natural interpretation under the dual descriptions associated with Langlands-dual groups.

We supplement this paper with an AI-assisted non-Abelian Hodge
audit companion developed for this work.\footnote{\url{https://github.com/ammohseni/nah-audit-companion}}  During
manuscript finalization, after the authors had developed the
scientific results and arguments presented here, the companion
provided a separate cross-check of the mathematical details
underlying our applications of non-Abelian Hodge theory against primary
mathematical sources
\cite{simpson1990harmonic,Simpson1992HiggsBA,Corlette1988flat,
Mochizuki2002,Mochizuki2003Asymptotic,mochizuki2004kobayashi,
mochizuki2009kobayashi,deligne2006equations}.
Its role was confined to these checks. 
The companion maintains permanent, reusable mathematical
checkpoints and provides separate mathematical and physics-facing
reports. Further details are given in the acknowledgments.

The paper is organized as follows. The introduction concludes with a
summary of our approach and a statement of our proposal.
Section~\ref{sec: class S material} reviews the class-\(\mathcal S\)
setup, plumbing geometry, and associated Hitchin-system data.
Section~\ref{SEC: NAH intro} introduces the non-Abelian Hodge framework,
while Section~\ref{sec:nah-asymptotics} develops the local monodromy and
flat-section asymptotics for one- and multi-parameter degenerations of a generic Riemann surface.
Section~\ref{sec:NAH for class S} specializes to
class-$\mathcal{S}$, presents the decorated cusp label and its
monodromy centralizer, and discusses physical observables.
Section~\ref{sec:examples} illustrates the construction using the
four-punctured sphere and the two-punctured torus, and concludes
with a discussion of a particular class of monodromy labels for
which the weak gauge algebra lies in the corresponding compact
reductive monodromy centralizer.
Finally, Section~\ref{sec: conclusion} summarizes our conclusions and
outlook.

\subsubsection*{Summary of our approach and proposal}

In this work, we develop a local non-Abelian Hodge-theoretic
framework that refines the description of weak-coupling cusps
in type-\(A\) class-\(\mathcal S\) theories. Under the appropriate
assumptions, we use non-Abelian Hodge theory to assign local monodromy
data of Hitchin--Simpson flat connections to the degenerating tubes,
and restrict to families admitting regular-singular logarithmic tube
models. We combine these decorations with the Picard--Lefschetz
monodromy of the UV curve. In type \(A\), we organize their discrete
types by semisimple eigenspace multiplicities and nilpotent Jordan
types.

Near a boundary divisor \(D_i=\{q_i=0\}\) in the conformal-manifold base,
the UV-curve fibration is locally described by the plumbing equation
\(u_i v_i=q_i\). As \(q_i\to 0\), the fiber develops an infinitely long
tube. The corresponding loop around \(D_i\) in the conformal-manifold
base changes the gluing angle of this tube and acts on the UV curve as
a Dehn twist. We obtain the local non-Abelian monodromy from the
holonomy of the Hitchin--Simpson flat connection around the core loop
of the tube. We choose a regular-singular logarithmic tube model and
write the local monodromy in its Jordan decomposition as
\be
    M_i=M_{s,i}M_{u,i}\ .
\ee
We define \(e_i=1/2\pi i\,\log M_{u,i}\) and choose a semisimple
logarithmic exponent \(\alpha_i\) of \(M_{s,i}\) that commutes with
\(e_i\), so that
\be
    R_i=\alpha_i+e_i,
    \qquad
    [\alpha_i,e_i]=0,
    \qquad
    M_i=\exp\bigl(2\pi iR_i\bigr).
\ee
The chosen semisimple exponent \(\alpha_i\) controls the power-law behavior
of flat sections and their transfer matrix, while \(e_i\) controls their
possible logarithmic growth. The pair \((M_{s,i},e_i)\) determines \(M_i\)
through \(M_{u,i}=\exp(2\pi i e_i)\). We use this pair, up to simultaneous
conjugation, as the non-Abelian Hodge monodromy decoration of the tube.

We define \(\mathfrak c_i\) to be a compact real form of the
reductive quotient of the Lie algebra centralizing \(M_i\),
and denote it by \(\mathfrak c_i(M_{s,i},e_i)\). The algebra \(\mathfrak c_i\) records the reductive continuous
symmetry preserving the local flat tube holonomy.

Alongside this local non-Abelian holonomy, the degenerating UV curve
carries Picard--Lefschetz monodromy. For a separating degeneration this
monodromy acts trivially on the compact weight-one cohomology of the
curve, whereas for a non-separating degeneration it is non-trivial and
has a nilpotent logarithm \(N^{(i)}_{\rm geom}\) satisfying
\(\bigl(N^{(i)}_{\rm geom}\bigr)^2=0\).

These two monodromies provide complementary data at a cusp:
the non-Abelian Hodge holonomy of the chosen Hitchin--Simpson flat connection
and the Abelian Picard--Lefschetz monodromy of the UV curve. We treat these
data together in a unified description of monodromy-decorated cusps. This allows us to track both the tube-wise symmetries preserving
the local holonomies and the asymptotic growth of flat sections
through the corresponding tube-transfer matrices.

\paragraph{Proposal: decorated cusp label.}
Let \(D_I=\bigcap_{i\in I}D_i\), with \(D_i=\{q_i=0\}\), be the intersection
of a collection of boundary divisors of the class-\(\mathcal S\) conformal
manifold. We associate to \(D_I\) the combined geometric, non-Abelian Hodge,
and weak-coupling data in the decorated cusp label
\be
    \mathfrak D_I
    =
    \left(
    \{\epsilon_i\}_{i\in I},
    \{M_{s,i},e_i\}_{i\in I},
    \{N_{\geom}^{(i)}\}_{i\in I},
    \{\mathfrak h_i\}_{i\in I}
    \right),
    \label{eq:decorated-cusp-label}
\ee
where \(\epsilon_i=s\) or \(ns\) records whether the divisor \(D_i\) is
separating or non-separating. The non-Abelian label \((M_{s,i},e_i)\)
records the semisimple part of the monodromy and the nilpotent logarithm
of its unipotent part. For the chosen \(\alpha_i\) with
\(M_{s,i}=\exp(2\pi i\alpha_i)\), the leading tube-transfer matrix has
power-law factor \(q_i^{\alpha_i}\) and logarithmic factor
\(\exp(e_i\log q_i)\). The data \(N_{\geom}^{(i)}\) record the ordinary
Picard--Lefschetz monodromy of the degenerating UV curve.

We treat the weak gauge algebra \(\mathfrak h_i\) as a separate
physical label of the \(i\)-th tube, fixed independently by the
class-\(\mathcal S\) fixture and gluing data. The centralizer belongs
to the chosen non-Abelian Hodge decoration and should not be
identified with a physical puncture flavor algebra or an intrinsic
invariant of the Deligne--Mumford boundary. A particular class of
monodromy labels, selected by imposing
\(\mathfrak h_i\subseteq\mathfrak c_i(M_{s,i},e_i)\) separately on
each tube, is discussed at the end of Section~\ref{sec:examples}.

\section{ Class-\texorpdfstring{$\mathcal S$}{S} theories}
\label{sec: class S material}

Class-\(\mathcal S\) theories are four-dimensional \(\mathcal N=2\)
superconformal field theories obtained by compactifying the six-dimensional
\(\mathcal N=(2,0)\) theory on a punctured Riemann surface
\cite{Gaiotto:2009we}; see \cite{Tachikawa:2013kta} for a review.  This
Riemann surface is usually called the \emph{UV curve}.  We denote it by
\(C_{g,n}\), where \(g\) is its genus and \(n\) is the number of punctures.  The
punctures encode codimension-two defect data of the six-dimensional theory,
while the complex structure of \(C_{g,n}\) determines the exactly marginal
couplings of the resulting four-dimensional theory.

The conformal manifold of marginal couplings, denoted by
\(\mathcal M_{g,n}\), is identified with the complex-structure moduli space of
the UV curve.  For a stable punctured curve \(2g-2+n>0\), its complex dimension is
\be
    \dim_{\mathbb C}\mathcal M_{g,n}=3g-3+n .
\ee
Equivalently, the moduli space is obtained as the quotient
\be
    \mathcal M_{g,n}
    \simeq
    \mathcal T_{g,n}/\Gamma_{g,n},
\ee
where \(\mathcal T_{g,n}\) is Teichm\"uller space and
\(\Gamma_{g,n}\) denotes the subgroup of the mapping class group that
preserves the specified puncture data. We work with the \emph{Deligne--Mumford compactification} of the conformal manifold,
written schematically as
\be
    \overline{\mathcal M}_{g,n}
    =
    \mathcal M_{g,n}\cup \{\text{cusps}\}.
\ee
The cusps correspond to nodal degenerations of \(C_{g,n}\)
\cite{DeligneMumford1969,Knudsen1983}. Near such a cusp, a plumbing
coordinate \(q\) parametrizes the degeneration: for \(0<|q|\ll1\),
the curve is smooth and contains a long tube, which pinches to a
node at \(q=0\) \cite{HubbardKoch2013}.

In the Hitchin-system description of the theory, the UV curve is equipped
with a meromorphic Higgs field \(\Phi\).  The allowed singular behavior of
\(\Phi\) at the punctures encodes the six-dimensional defect data, while its
spectral curve gives the Seiberg--Witten curve of the four-dimensional theory
\cite{Hitchin1987StableBundles,Seiberg:1994rs,Donagi:1995cf}.  Thus the punctured curve
\(C_{g,n}\) carries both the conformal-manifold data through its complex
structure and the Hitchin data that organize the Coulomb-branch geometry.

In this work we focus on type-\(A_{N-1}\) class-\(\mathcal S\) theories.  These
are obtained by placing the six-dimensional \(\mathcal N=(2,0)\) theory of
type
$\g=\mathfrak{sl}_N$
on the punctured Riemann surface \(C_{g,n}\).  From the six-dimensional point
of view, placing the \(\mathcal N=(2,0)\) theory directly on
\(\mathbb R^{1,3}\times C_{g,n}\) would generically break supersymmetry.
Supersymmetry is preserved by performing a partial topological twist along
\(C_{g,n}\).
A useful string-theoretic realization is provided by M-theory.  The
type-\(A_{N-1}\) theory arises as the low-energy theory on a stack of \(N\)
coincident M5-branes.  Class-\(\mathcal S\) theories are obtained by wrapping
these M5-branes on \(C_{g,n}\), with the partial twist ensuring that
supersymmetry is preserved in the four noncompact directions.  The punctures
of \(C_{g,n}\) correspond to codimension-two defects associated with the
M5-brane configuration.

\subsection{Illustrative example: The four-punctured sphere} \label{sec:Illustrative_example}

In this section, we use the four-punctured sphere as an illustrative example to describe the UV curve, its conformal manifold, the relation between them, and their degenerations. Take
$g=0,\ n=4$, 
where the UV curve is a four-punctured sphere.  In this case, the conformal manifold is complex 1-dimensional, while each UV curve is a complex curve. The nodal
degenerations are separating degenerations produced by the collision of
punctures.

Let the UV curve be a sphere $\mathbb P^1_z$ with four punctures at
$z_1,z_2,z_3,z_4$.  We define the cross-ratio coordinate
\be
    \lambda
    =
    \frac{(z_1-z_3)(z_2-z_4)}
         {(z_1-z_4)(z_2-z_3)}\ .
\ee
Using the $PSL(2,\mathbb{C})$ automorphism of the sphere, we fix three punctures by
the convention
\be
    z_3\to0\,,\qquad
    z_2\to1\,,\qquad
    z_4\to\infty\ ,
\ee
so that $z_1\to\lambda$.

The conformal manifold is the complex-structure moduli space of the UV curve
and is parametrized by \(\lambda\).  Explicitly \footnote{If some punctures carry identical defect data, the physical conformal manifold
may involve a further quotient by permutations preserving the specified puncture data,
under which some of these cusps can be identified.}
\be
    \mathcal M_{0,4}
    \cong
    \mathbb P^1_\lambda\setminus\{0,1,\infty\}
    \cong
    \mathbb H/\Gamma(2).
\ee
The corresponding quotient map is the modular lambda function
\be
    \lambda(\tau_\lambda)
    =
    \frac{\vartheta_2(\tau_\lambda)^4}
         {\vartheta_3(\tau_\lambda)^4},
\ee
which is invariant under \(\Gamma(2)\) and identifies
\(\mathbb H/\Gamma(2)\) with
\(\mathbb P^1_\lambda\setminus\{0,1,\infty\}\).
Here \(\tau_\lambda\) is the modular parameter entering the lambda
function and is distinct from the local weak-coupling coordinates
introduced below. The three points \(\lambda=~0,1,\infty\) are cusps.  They correspond to the
three possible pairwise collisions of punctures. We will be interested in the behavior of the conformal manifold near each of
the three cusps, which correspond to the three  pairwise degenerations of the four-punctured sphere.

We can describe the universal UV curve as
\be
    \mathcal C=\bigcup_{\lambda}C_\lambda ,
\ee
where $\mathcal C$ is obtained from $\mathbb{CP}^2$, with homogeneous
coordinates $(x,y,z)\sim c(x,y,z)$, by blowing up the four points
\be
    E_1=(1,0,0),\qquad
    E_2=(0,1,0),\qquad
    E_3=(0,0,1),\qquad
    E_4=(1,1,1).
\ee
The projection
\(
    \pi:\mathcal C\longrightarrow\overline{\mathcal M}_{0,4}
\)
is described by
\be
    \lambda\,x(y-z)+y(z-x)=0 .
    \label{eq:universal-curve}
\ee

Equivalently, the total space \(\mathcal C\) is fibered over
\(\mathcal M_{0,4}\), with fiber
\(
C_\lambda=\pi^{-1}(\lambda).
\)
 \(\mathcal M_{0,4}\) is the base of the fibration, while the UV
curves \(C_\lambda\) are the fibers. Schematically,
\be
\begin{array}{c}
C_\lambda \subset \mathcal C \\
\downarrow \\
\lambda \in \mathcal M_{0,4}\ .
\end{array}
\ee
The fiber $C_\lambda=\pi^{-1}(\lambda)$ is smooth at a generic point, but
degenerates at $\lambda=0,1,\infty$ as
\be
C_0=\{y(z-x)=0\}\,,\qquad C_1=\{z(x-y)=0\}\,,\qquad C_\infty=\{x(y-z)=0\}\ .
\ee
These are singular fibers of the universal UV curve and cusps of the
conformal manifold.

\subsection{Plumbing geometry of the weak-coupling cusp}

For the four-punctured sphere, the family of UV curves is
\be
    C_\lambda:\qquad
    \lambda\,x(y-z)+y(z-x)=0 .
    \label{eq:universal-curve-local}
\ee
The coordinate \(\lambda\) is a global coordinate on
\(\mathcal M_{0,4}\), but it is not the most convenient local coordinate at
all three cusps.  Instead, near each cusp we introduce a \emph{local plumbing
coordinate} \(q\), which vanishes at that cusp.  Concretely, we use
\be
    q_{\lambda=0}=\lambda ,
    \qquad
    q_{\lambda=1}=1-\lambda ,
    \qquad
    q_{\lambda=\infty}=\frac{1}{\lambda} .
    \label{eq:three-local-q-coordinates}
\ee
These describe the cusps at $\lambda=0,1,\infty$ respectively. 

The three cusps correspond to the three possible separating degenerations of
the four-punctured sphere:
\be
\begin{array}{c|c|c|c}
\text{cusp}
&
\text{\ local coordinate\ }
&
\text{\ colliding punctures\ }
&
\text{singular fiber}
\\[1mm]
\hline
\lambda=0
&
q_{\lambda=0}=\lambda
&
z_1\to z_3
&
C_0=\{y=0\}\cup\{z=x\}
\\[1mm]
\lambda=1
&
q_{\lambda=1}=1-\lambda
&
z_1\to z_2
&
C_1=\{z=0\}\cup\{x=y\}
\\[1mm]
\lambda=\infty
&
q_{\lambda=\infty}=1/\lambda
&
z_1\to z_4
&
C_\infty=\{x=0\}\cup\{y=z\}
\end{array}
\ee
In each case, the singular fiber consists of two components meeting at a
node.  The plumbing description replaces the nodal equation
$uv=0$,
by the smoothed equation
\be
    uv=q\ .
    \label{eq:local-plumbing-q}
\ee
The variables \(u\) and \(v\) are local coordinates on the two branches of
the nodal curve. The coordinate \(q\) should be understood as \(q_{\lambda=0}\), \(q_{\lambda=1}\), or
\(q_{\lambda=\infty}\), depending on the weak-coupling frame.  

Let us make this explicit at the cusp \(\lambda=0\).  Near the node, we work
in the patch \(x=1\) and define
$b=z-1$.
Then the curve equation becomes
\be
    yb+\lambda y-\lambda b-\lambda=0 .
    \label{eq:curve-equation-local-zero}
\ee
Introducing
\be
    u=y-\lambda,
    \qquad
    v=b+\lambda ,
\ee
one finds
\be
    uv
    =
    (y-\lambda)(b+\lambda)
    =
    \lambda(1-\lambda) .
\ee
  Thus, for \(q=\lambda(1-\lambda)\sim\lambda\), the local model is
\(
    uv=q
\). The same construction applies at the other two cusps.

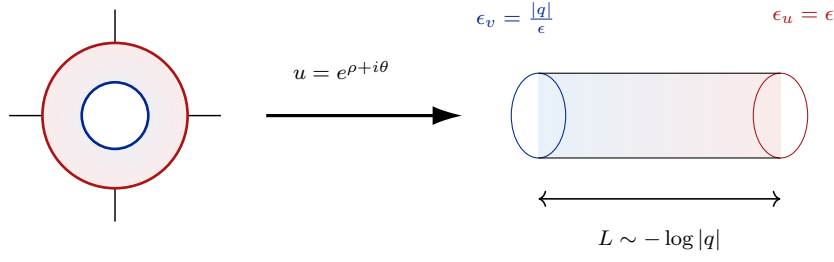
\begin{figure}
\begin{center}
\begin{tikzpicture}[x=2cm,y=2cm]

\coordinate (A0) at (1.7,0.3);

\draw[line width=0.55pt] ($(A0)+(-0.7,0)$) -- ($(A0)+(0.7,0)$);
\draw[line width=0.55pt] ($(A0)+(0,-0.7)$) -- ($(A0)+(0,0.7)$);

\shade[inner color=lightublue, outer color=lightvred] (A0) circle (0.48);

\fill[white] (A0) circle (0.22);

\draw[vbranch] (A0) circle (0.48);   % outer circle red
\draw[ubranch] (A0) circle (0.22);   % inner circle blue

\node[hepblue, font=\scriptsize] at ($(A0)+(0,0.78)$) {};
\node[branchred, font=\scriptsize] at ($(A0)+(0,-0.82)$) {};

\draw[mainarrow] (2.7,0.3) -- (4.0,0.3);
\node[font=\scriptsize] at (3.2,0.6) {$u=e^{\rho+i\theta}$};

\shade[left color=lightublue,right color=lightvred]
  (4.5,0.02) rectangle (6.1,0.58);

\draw[hepblue]   (4.5,0.3) ellipse (0.18 and 0.28);
\draw[branchred] (6.1,0.3) ellipse (0.18 and 0.28);
\draw (4.5,0.58) -- (6.1,0.58);
\draw (4.5,0.02) -- (6.1,0.02);

\node[hepblue, font=\scriptsize] at (4.35,0.95) {$\epsilon_v=\frac{|q|}{\epsilon}$};
\node[branchred, font=\scriptsize] at (6.25,0.95) {$\epsilon_u=\epsilon$};

\draw[<->, line width=0.6pt] (4.5,-0.25) -- (6.1,-0.25);
\node[font=\scriptsize] at (5.3,-0.52) {$L\sim-\log|q|$};

\end{tikzpicture}
\end{center}
\caption{The node plumbed into a long tube. Here \(\epsilon_u=\epsilon\) and \(\epsilon_v=\frac{|q|}{\epsilon}\) are the cutoff radii.}
\label{tube}
\end{figure}

For \(0<|q|\ll1\), the node is replaced by a smooth annulus. 
Using \eqref{eq:local-plumbing-q} and imposing the cutoffs \(|u|<\epsilon\) and \(|v|<\epsilon\), the annulus is
\be
    \frac{|q|}{\epsilon}<|u|<\epsilon .
\ee
Introducing the cylindrical coordinate
$u=e^{\rho+i\theta}$, with $    \theta\sim\theta+2\pi$,
the annulus becomes a cylinder
\be
    \log\left(\frac{|q|}{\epsilon}\right)
    <
    \rho
    <
    \log\epsilon .
\ee
Its length in the \(\rho\)-direction is
\be
    L
    =
    -\log|q|+2\log\epsilon
    \sim
    -\log|q| .
\ee
The cutoff-dependent term \(2\log\epsilon\) does not affect the leading cusp
behavior.  Thus every weak-coupling cusp of the four-punctured sphere is
described by a long tube (see Figure~\ref{tube}).
In the weak-coupling duality frame associated with the chosen cusp, the
nodal limit \(q\to0\) corresponds to the weak-coupling limit of the gauge
group associated with the tube.

After choosing a cusp, we denote the nearby fiber by \(C_q\).  This
notation emphasizes that \(q\) is a local coordinate on the conformal
manifold adapted to the chosen weak-coupling degeneration.  Thus \(C_q\)
means the fiber expressed in the appropriate local chart for the cusp under
consideration.  The limiting fiber \(C_{q=0}\) is nodal, while for
\(0<|q|\ll1\), the fiber \(C_q\) is smooth and contains a long tube
described locally by
\(
    uv=q 
\).

\subsection{Hitchin-system data for the UV curve}
\label{subsec:hitchin-system-data}

We now recall the Hitchin-system data associated with the UV curve of Type A theory.  This
sets the stage for the non-Abelian Hodge description.  For a
fixed smooth fiber \(C_\lambda\), the relevant data are a holomorphic
\(SL(N,\mathbb C)\) bundle \(V\to C_\lambda\) and a meromorphic Higgs field
\be
    \Phi
    \in
    H^0\!\left(
        C_\lambda,
        K_{C_\lambda}(D)\otimes \End_0(V)
    \right),
    \qquad
    D=\sum_a p_a ,
\ee
where the points \(p_a\) are the punctures.  The subscript in
\(\End_0(V)\) implements the type-\(A_{N-1}\) condition
\(\Tr\Phi=0\) (see Appendix \ref{app:higgs-bundles}).  Thus a Higgs bundle is the pair
$(V,\Phi)$.
Locally, the Higgs field can be written as
\be
    \Phi=\varphi(z)\,dz,
    \qquad
    \varphi(z)\in \mathfrak{sl}_N .
\ee
The meromorphic singularities of \(\Phi\) at the punctures encode the
codimension-two defect data of the class-\(\mathcal S\) construction.

The Coulomb-branch geometry is encoded in the \emph{Seiberg--Witten curve}
\be
    \Sigma_\lambda\subset T^*C_\lambda .
\ee
Let \(w\) denote the canonical one-form on \(T^*C_\lambda\).  The
Seiberg--Witten curve is obtained from the spectral equation
\be
    \Sigma_\lambda:\qquad
    \det\bigl(w-\Phi\bigr)=0 .
\ee
Equivalently, in a local coordinate with \(w=x\,dz\), this equation takes the
form
\be
    x^N+\sum_{k=2}^{N}\phi_k(z)\,x^{N-k}=0,
    \label{eq:SW-spectral-curve}
\ee
where \(x\) is the local spectral coordinate.  The absence of the
\(x^{N-1}\) term follows from \(\Tr\Phi=0\).  The coefficients \(\phi_k\) are
meromorphic \(k\)-differentials on \(C_\lambda\).

The punctures of \(C_\lambda\) prescribe the allowed singular behavior of
\(\Phi\). Near a regular puncture at \(z=0\), one has
\be
    \Phi(z)
    \sim
    \operatorname{Res}_{z=0}(\Phi)\,\frac{dz}{z}
    +\cdots .
    \label{eq:Higgs-field-residue}
\ee
At the superconformal point, the residue at a regular puncture is
nilpotent. Its conjugacy class,
\(
    \mathcal O_a\subset\mathfrak g 
\),
is called the Hitchin nilpotent orbit associated with the puncture. The same puncture also admits an equivalent label, called the Nahm
partition. In type \(A\), the Hitchin partition is the transpose of the Nahm partition.  In the Hitchin-system language, \(\Phi\) is the Hitchin Higgs field and \(\Sigma_\lambda\) is its spectral curve. Under the appropriate stability
and regularity assumptions, non-Abelian Hodge theory relates these
Higgs-bundle data to a Hitchin--Simpson flat connection.  In what follows, fiberwise flat connections, especially their
holonomy around the long tube that appears when \(C_\lambda\)
degenerates, will be used to organize the local monodromy and the
tube-wise monodromy centralizers associated with the chosen tube
data near cusps of the conformal manifold.

\section{Non-Abelian Hodge theory: general principles}\label{SEC: NAH intro}

We first recall the non-Abelian Hodge-theoretic framework used in the rest of
the paper.  Let \(X\) be a punctured Riemann surface, and let
\(G_{\mathbb C}\) be a complex reductive Lie group.  Non-Abelian Hodge theory relates
three descriptions of the same geometric data: flat
\(G_{\mathbb C}\)-bundles, monodromy representations of \(\pi_1(X)\), and
Higgs bundles on \(X\).  In this paper we use this correspondence in its
associated vector-bundle form. In the class-\(\mathcal S\) applications below, we specialize to the
type-\(A\) case \(G_{\mathbb C}=SL(N,\mathbb C)\), use its defining
representation, and take \(X\) to be the UV curve in the corresponding family. The asymptotic mechanism used later is developed
in the next section. Readers primarily interested in the application may skip
this section.

\subsection{Flat connections from Higgs bundles}

For comparison with ordinary Abelian Hodge theory, first take \(X\) to be
compact. One then studies the cohomology group
\be
    H^1(X,\mathbb C)
    \simeq
    \operatorname{Hom}(H_1(X,\mathbb Z),\mathbb C),
\ee
together with its Hodge decomposition. Equivalently, since
\(H_1(X,\mathbb Z)\) is the Abelianization of \(\pi_1(X)\), ordinary Hodge
theory describes the Abelian monodromy problem as a vector space.

The non-Abelian Hodge theory
\cite{Donaldson1987,Hitchin1987SelfDuality,Corlette1988flat,Simpson1992HiggsBA}
provides the corresponding non-Abelian description in terms of the character variety
\be
    \M_{\rm Betti}(X,G_{\mathbb C})
    =
    \operatorname{Hom}(\pi_1(X),G_{\mathbb C})/G_{\mathbb C}.
\ee
A point of the Betti space corresponds to a representation
\be
    \rho:\pi_1(X)\longrightarrow G_{\mathbb C},
\ee
with two representations identified if they differ by an overall
conjugation by \(G_{\mathbb C}\). This identification is the meaning
of the quotient in the definition of \(\mathcal M_{\mathrm{Betti}}\).\footnote{More precisely, on the Betti side one takes
simultaneous conjugacy classes, keeping the closed or polystable orbits.  This
quotient is often denoted by \( /\!\!/ \).}

With the complex structure on \(X\) fixed, the same data may be
described in Dolbeault language by a \(G_{\mathbb C}\)-Higgs bundle.
The corresponding moduli space is \footnote{In the Dolbeault and de Rham descriptions, \(\sim\) identifies
equivalent Higgs bundles or flat connections, respectively.}
\be
   \M_{\rm Dol}(X,G_{\mathbb C})
=
\{G_{\mathbb C}\text{-Higgs bundles}\}/\!\sim.
\ee
 In a finite-dimensional representation of \(G_{\mathbb C}\), we denote the
corresponding Higgs bundle by \((V,\Phi)\), where \(V\to X\) is the associated
holomorphic vector bundle. The condition
\be
    \bar\partial_V\Phi=0
\ee
means that \(\Phi\) is holomorphic with respect to the holomorphic structure
on \(V\).

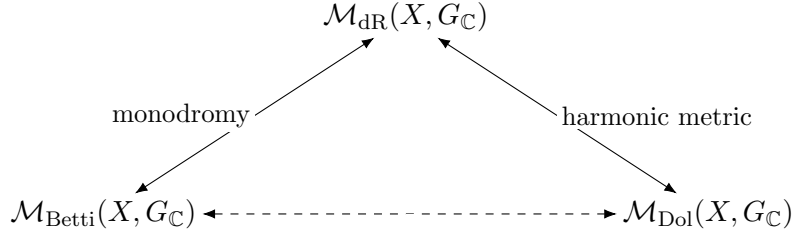
\begin{figure}[H]
\centering
\begin{tikzpicture}[
    >=Latex,
    every node/.style={align=center},
    mod/.style={inner sep=2pt}
]

\node[mod] (dR) at (0,2.6)
    {$\mathcal M_{\mathrm{dR}}(X,G_{\mathbb C})$};

\node[mod] (B) at (-4.0,0)
    {$\mathcal M_{\mathrm{Betti}}(X,G_{\mathbb C})$};

\node[mod] (D) at (4.0,0)
    {$\mathcal M_{\mathrm{Dol}}(X,G_{\mathbb C})$};

\draw[<->] (dR) -- node[left, fill=white, inner sep=1.5pt]
    {\small monodromy} (B);

\draw[<->] (dR) -- node[right, fill=white, inner sep=1.5pt]
    {\small harmonic metric} (D);

\draw[<->, dashed] (B) -- node[below, fill=white, inner sep=1.5pt]
    {} (D);

\end{tikzpicture}
\caption{Schematic relation among the Betti, de Rham, and Dolbeault
descriptions.}
\label{fig:nah_triangle}
\end{figure}

The bridge between the Betti and Dolbeault descriptions is provided
by a flat connection, or equivalently by a point in the de Rham
moduli space
\be
   \M_{\rm dR}(X,G_{\mathbb C})
=
\{\text{flat }G_{\mathbb C}\text{-connections}\}/\!\sim.
\ee
This flat connection is constructed using a harmonic metric
\(h\) satisfying the Hitchin--Simpson equations. On the de Rham side, the
same data are described by a flat bundle
$(V_{\mathrm{dR}},\nabla)$,
where \(V_{\mathrm{dR}}\to X\) is the holomorphic vector bundle associated
with the representation of \(G_{\mathbb C}\), and \(\nabla\) is the
induced holomorphic connection, with holonomy in \(G_{\mathbb C}\), satisfying
the flatness condition
\be
    \nabla^2=F_\nabla=0.
\ee
Its parallel transport determines the monodromy representation
\(
    \rho:\pi_1(X)\longrightarrow G_{\mathbb C}
\).
The relations among these three descriptions are summarized
schematically in Figure~\ref{fig:nah_triangle}.
For punctured curves, this correspondence is formulated in the tame
filtered, or parabolic, setting
\cite{simpson1990harmonic,simpson1996hodge,mochizuki2004kobayashi},
with the appropriate stability, degree, and regularity conditions imposed at
the punctures.  In the general
\(G_{\mathbb C}\)-bundle setting, one must specify additional local structure
at the punctures; in this paper, however, we restrict to the associated
vector-bundle framework. For the nilpotent, trivial-parabolic case, we use
Mochizuki's treatment \cite{Mochizuki2002} only within that restricted
setting. We will be interested in the local consequences of the correspondence:
the construction of the Hitchin--Simpson flat connection from a Higgs bundle
and the asymptotic behavior of its flat sections near a puncture or a
degenerating tube.  These asymptotic consequences are discussed in the next
section.

Here is an outline of how Higgs-bundle data give rise to a flat connection
and its monodromy representation. To pass from a Higgs bundle to a flat connection, one chooses a Hermitian
metric \(h\) on \(V\).  The metric determines a unique connection \(D_h\) on
\(V\), called the \emph{Chern connection},
\be
    D_h=\bar\partial_V+\partial_h,
\ee
which is compatible with both the holomorphic structure and the Hermitian
metric.  Here \(\bar\partial_V\) is the \((0,1)\) part of the connection and
defines the holomorphic structure on \(V\), while \(\partial_h\) is the
\((1,0)\) part determined by
\be
    D_h h=0 .
\ee
In a local holomorphic frame, if \(H(z,\bar z)\) is the Hermitian matrix
representing \(h\), then
\be
    \bar\partial_V=\bar\partial,
    \qquad
    \partial_h=\partial+H^{-1}\partial H,
\ee
and therefore
\be
    D_h=d+ H^{-1}\partial H .
\ee
The connection \(D_h\) carries the same gauge indices as the Higgs field.

The harmonic metric is selected by the Hitchin--Simpson equation
\be
    F_{D_h}+[\Phi,\Phi^{\dagger_h}]=0.
\ee
The collection
\(
    (V,\bar\partial_V,\Phi,h)
\)
is called a \emph{harmonic bundle}.  
The harmonic metric determines the \emph{Hitchin--Simpson connection}
\be
    \nabla_h
    =
    D_h+\Phi+\Phi^{\dagger_h}.
\ee
The Hitchin--Simpson equation implies that this connection is flat,
\be
    F_{\nabla_h}=0.
\ee
Thus, under the stability and regularity hypotheses which supply the harmonic
metric, the Higgs-bundle data determine a flat connection and hence a monodromy
representation of \(\pi_1(X)\), up to the usual gauge equivalences.

\paragraph{Local monodromy.}
The flat connection gives a representation
\be
    \rho:\pi_1(X)\to G_\mathbb{C} .
\ee
Let the Hitchin--Simpson flat connection be denoted by
\be
    \nabla_h=d+ A_h,
\ee
and let \(\gamma\) be a closed loop on \(X\).  The monodromy representation on
\(X\) assigns to \(\gamma\) the parallel-transport operator
\be
    M(\gamma)
    =
    \rho(\gamma)
    =
    P\exp\left(-\oint_\gamma A_h\right)
    \in G_{\mathbb C}.
    \label{eq:monodromy}
\ee
Because the Betti moduli space is quotiented by conjugation, two
representations related by
\be
    \rho(\gamma)\sim g\,\rho(\gamma)\,g^{-1},
    \qquad
    g\in G_{\mathbb C},
\ee
represent the same point.  Hence a monodromy matrix is meaningful only up to
conjugacy.

It is important to distinguish the residue of the Higgs field from the
monodromy of the flat connection.  Even if the Higgs field has a nilpotent
residue at a puncture, the monodromy of \(\nabla_h\) is not determined by that
nilpotent residue alone.  It also receives contributions from the Chern
connection \(D_h\) and from the adjoint Higgs field \(\Phi_h^\dagger\).  In the
local analysis below, \(R\) denotes the logarithmic exponent of the resulting
flat connection, not the residue of \(\Phi\) alone.

\subsection{The Abelian limit}

The preceding non-Abelian Hodge correspondence reduces in the Abelian limit
to the familiar cohomological description. For this illustration, take
\(X\) to be compact and work without logarithmic singularities. Set
\be
    G_{\mathbb C}=\mathbb C^\ast .
\ee
Then a representation of \(\pi_1(X)\) into \(\mathbb C^\ast\) only depends on
the homology class of a loop:
\be
    \operatorname{Hom}(\pi_1(X),\mathbb C^\ast)
    =
    \operatorname{Hom}(H_1(X,\mathbb Z),\mathbb C^\ast).
\ee
This is because the homology group \(H_1(X,\mathbb Z)\) is the Abelianization
of the fundamental group.  Passing to the tangent space, or equivalently to the
space of infinitesimal deformations, gives
\be
    T\operatorname{Hom}(\pi_1(X),\mathbb C^\ast)
    =
    \operatorname{Hom}(H_1(X,\mathbb Z),\mathbb C)
    =
    H^1(X,\mathbb C).
\ee
Thus, in the Abelian limit, the Betti moduli space linearizes to the usual
cohomology group \(H^1(X,\mathbb C)\).

On the Dolbeault side, a Higgs bundle reduces to a pair \((L,\Phi)\), where
\(L\) is a line bundle and \(\Phi\) is a one-form.  Variations of the
holomorphic structure of \(L\) give \((0,1)\)-forms, while variations of
\(\Phi\) give \((1,0)\)-forms.  Thus locally
\be
    T\M_{\rm Dol}
    =
    H^{0,1}(X)\oplus H^{1,0}(X),
\ee
which is precisely the Hodge decomposition of \(H^1(X,\mathbb C)\).

On the de Rham side, an Abelian flat connection has the form
\be
    \nabla=d+\alpha,
\ee
and flatness simply means
\be
    d\alpha=0 .
\ee
Therefore the Abelian de Rham moduli space is described by closed one-forms,
modulo gauge equivalence.  This again gives the usual cohomological
description.  The non-Abelian theory is a nonlinear version of this story: the
linear vector space \(H^1(X,\mathbb C)\) is replaced by a character variety,
and the Hodge decomposition is replaced by the correspondence between flat
connections and Higgs bundles.

\subsection{Logarithmic singularities}

For the applications below, we use the standard local model for a tame
harmonic bundle
\cite{deligne2006equations,simpson1990harmonic,mochizuki2009kobayashi}.
Near a marked point or a degeneration, its de Rham connection is regular
singular and can be represented by a logarithmic connection after choosing an
appropriate extension.  Let
\(X\) be a punctured Riemann surface, written as
\[
    X=\overline X\setminus D ,
\]
where \(\overline X\) is a smooth compact Riemann surface and
\(D\subset \overline X\) is the divisor of punctures.  Let
\((V_{\mathrm{dR}},\nabla)\) be the \emph{flat holomorphic vector bundle}\footnote{This is a holomorphic vector bundle equipped with a flat connection, obtained from the corresponding harmonic bundle.} associated,
in the chosen finite-dimensional representation, with the flat
\(G_{\mathbb C}\)-bundle on \(X\).  A logarithmic extension across \(D\) is a
holomorphic vector bundle
\(\overline V_{\mathrm{dR}}\to\overline X\) extending
\(V_{\mathrm{dR}}\), equipped with a connection
\be
    \nabla:
    \overline V_{\mathrm{dR}}
    \longrightarrow
    \overline V_{\mathrm{dR}}
    \otimes \Omega^1_{\overline X}(\log D) .
\ee
The connection one-form is
\(\mathfrak g_{\mathbb C}\)-valued in the chosen representation, and
\(\nabla\) is allowed to have logarithmic poles along \(D\).

If \(p\in D\) and \(\zeta\) is a local coordinate with
\(D=\{\zeta=0\}\), then \(\Omega^1_{\overline X}(\log D)\) is locally
generated by \(d\zeta/\zeta\).  The leading regular-singular part of the
connection may therefore be written as
\be
    \nabla
    \simeq
    d-R_p\frac{d\zeta}{\zeta},
    \qquad
    R_p\in
    \mathfrak g_{\mathbb C}.
\ee
We choose a logarithmic model, with a fixed logarithmic
branch, so that \(R_p\) is the corresponding logarithmic exponent of
the local monodromy. For this choice, the exponent and local monodromy
are defined up to \(G_{\mathbb C}\)-conjugacy, with
\be
    M_p=\exp(2\pi iR_p)\in G_{\mathbb C}.
\ee

In this setting, the tame condition means, roughly, that flat sections have
controlled polynomial, or power-law, growth near the puncture as
\(\zeta\to0\).  In the non-Abelian Hodge correspondence, this controlled growth
is encoded by a parabolic, or filtered, extension of the bundle across the
special point.  The detailed construction of this filtered extension is part of
the tame non-Abelian Hodge correspondence.

\subsection{Non-Abelian monodromy and growth}
 Let \(\zeta\) be a local coordinate
centered at the special point of a curve \(X\), so that the point lies at
\(\zeta=0\).  For a chosen tame harmonic bundle whose de Rham connection admits
a logarithmic extension at \(\zeta=0\), the Hitchin--Simpson flat connection has leading regular-singular model
\be
    \nabla_h
    \sim
    d-R\,\frac{d\zeta}{\zeta},
    \label{eq:flat-connection-singularity}
\ee
where \(R\in\mathfrak g_{\mathbb C}\), up to conjugacy and the choice of logarithmic branch.  Using
\eqref{eq:monodromy}, the local monodromy around \(\zeta=0\) is
\be
    M=\exp(2\pi i R).
\ee

A \emph{flat section} is a section \(s\) of the vector bundle satisfying
\be
    \nabla_h s=0 .
\ee
For the local connection~\eqref{eq:flat-connection-singularity}, the flat-section
equation and its solution are
\begin{align}
    &\frac{d s(\zeta)}{d\zeta}
    =\frac{R}{\zeta}s(\zeta),
    \label{eq:local-flat-section-equation}\\
    &s(\zeta)
    =\zeta^R s_0
      =\exp(R\log\zeta)s_0 .
    \label{eq:local-flat-section-solution}
\end{align}
Analytic continuation around \(\zeta=0\) sends
\(\zeta\mapsto\zeta e^{2\pi i}\), and therefore \(s\mapsto Ms\).

Choose a logarithm of the monodromy and decompose \(R\) into commuting
semisimple and nilpotent parts:
\be
    R=\alpha+e,
    \qquad
    [\alpha,e]=0 .
\ee
Equivalently, the Jordan decomposition of the monodromy is
\be
    M=M_s M_u,
    \qquad
    M_s=\exp(2\pi i\alpha),
    \qquad
    M_u=\exp(2\pi i e),
\ee
where \(M_s\) is semisimple and \(M_u\) is unipotent.  Since \(e\) is nilpotent,
there exists an integer \(\ell\) such that
\be
    e^\ell=0,
    \qquad
    e^{\ell-1}\neq0 .
\ee
Using \([\alpha,e]=0\), the flat section may be written as the following
local growth formula:
\be
\begin{aligned}
    s(\zeta)
    &=
    \zeta^\alpha \exp(e\log\zeta)s_0
    \\
    &=
    \zeta^\alpha
    \left(
        I+\sum_{n=1}^{\ell-1}
        e^n\frac{(\log\zeta)^n}{n!}
    \right)s_0 .
\end{aligned}
    \label{eq:leading-flat-section-growth}
\ee
This formula describes the leading logarithmic model; a rigorous analytic
treatment of flat-section growth for tame harmonic bundles requires the
standard hypotheses on the behavior at the puncture. The nilpotent
trivial-parabolic case is treated by Mochizuki, while the general tame local
asymptotic estimates are supplied by Mochizuki's later work
\cite{simpson1990harmonic,Mochizuki2002,Mochizuki2003Asymptotic,
mochizuki2004kobayashi,mochizuki2009kobayashi}.
For a vector \(s_0\) on which \(e^{\ell-1}s_0\neq0\), the highest logarithmic term is
\be
    s(\zeta)
    \sim
    \zeta^\alpha
    \frac{(\log\zeta)^{\ell-1}}{(\ell-1)!}\,e^{\ell-1}s_0 .
\ee
This is the highest logarithmic term within the generalized
\(\alpha\)-eigenspace containing \(s_0\). If several semisimple eigenvalues
occur, the globally dominant radial asymptotic is determined by comparing
the corresponding power-law factors. \emph{Thus, in the normalized logarithmic
model used here, the local growth has two distinct pieces.  The chosen
semisimple exponent \(\alpha\) records the selected power-law factor, while the
nilpotent part \(e\) gives the finite logarithmic polynomial.}  This is the
local non-Abelian asymptotic data associated with a logarithmic degeneration.

\paragraph{Abelian limit.}
In the Abelian case \(G_{\mathbb C}=\mathbb C^\ast\), a logarithmic flat
connection near \(\zeta=0\) has the form
\be
    \nabla
    \simeq
    d-r\,\frac{d\zeta}{\zeta},
    \qquad
    r\in\mathbb C .
\ee
The flat-section equation gives
\be
    s(\zeta)=\zeta^r s_0,
\ee
and the local monodromy is
\be
    M=e^{2\pi i r}\in\mathbb C^\ast .
\ee
Equivalently, this is the \(1\times1\) version of the above discussion:
\be
    R=\alpha=r,
    \qquad
    e=0 .
\ee
\emph{Thus power-law behavior is already present in the Abelian case,
while a nontrivial nilpotent part introduces logarithmic factors
governed by its discrete Jordan type.}

Globally, however, Abelian flat-bundle monodromy is not purely local.
A closed one-form can have nontrivial periods along the cycles of \(X\),
and these periods determine its monodromy around non-contractible loops.
The flat-bundle monodromy should be distinguished from the geometric,
or Picard--Lefschetz, monodromy of a degenerating family, which acts on
the homology or cohomology of the fibers. Nevertheless, both structures
can be attached to the same degeneration and should be treated together.

\section{Non-Abelian Hodge theory: asymptotics}
\label{sec:nah-asymptotics}

Using non-Abelian Hodge theory, we develop a local framework for describing
the monodromy data and leading asymptotic behavior associated with
degenerations of punctured Riemann surfaces equipped with Higgs-bundle data.
In the general discussion, we denote the Riemann surface by \(X\); upon
specializing to class-\(\mathcal S\) theories, this surface is the UV curve,
which we denote by \(C\). Near a boundary divisor, the flat-connection
description gives direct access to non-Abelian monodromy, which controls the
asymptotic growth of flat sections: its semisimple part gives power-law
behavior, while its nilpotent part gives logarithmic behavior. Meanwhile, the
degenerating curve carries Abelian geometric monodromy on its homology and
period data.

Throughout, we use the \emph{local logarithmic tube model}. In each plumbing
neighborhood, we choose a fiberwise Hitchin--Simpson flat connection with
regular-singular logarithmic behavior and denote its non-Abelian exponent by
\(R_i\). The complex-structure degeneration determines the tube geometry,
whereas \(R_i\) is additional asymptotic flat-connection data, defined up to
conjugacy and a choice of logarithmic branch. At a multi-divisor cusp, the
base-coordinate loops and the corresponding Dehn twists about disjoint
vanishing cycles commute, although the tube holonomies obtained by evaluating
the fiberwise flat connection need not. Transport through multiple tubes is
therefore described by an ordered composition of single-tube transfer
operators.

\subsection{One-parameter degenerations}
\label{subsec:one-variable-nah-degeneration}

 A one-parameter degeneration is described by a family of Riemann surfaces over
a local one-dimensional base,
\be
    \pi:\mathcal X\longrightarrow B,
    \qquad
    B\simeq \Delta ,
\ee
where \(\Delta\) is a small disk and \(q\) is a coordinate on \(B\).  The smooth
members of the family lie over the punctured disk
\(
    \Delta^\ast=\Delta\setminus\{0\}
\),
and we denote them by
\be
    X_q=\pi^{-1}(q),\qquad q\in \Delta^\ast .
\ee
In a nodal degeneration, the local geometry of \(\mathcal X\) near the node is
described by the plumbing equation
\be
    uv=q,
    \label{eq:generic-plumbing-t}
\ee
where \(q\) is the base coordinate.  The central member \(X_0\) has a node,
while for \(0<|q|\ll1\) this node is smoothed and replaced by a long tube.

\subsubsection{Boundary divisors and degeneration coordinates}

\begin{figure}[t]
    \centering
    \includegraphics[width=0.96\textwidth]{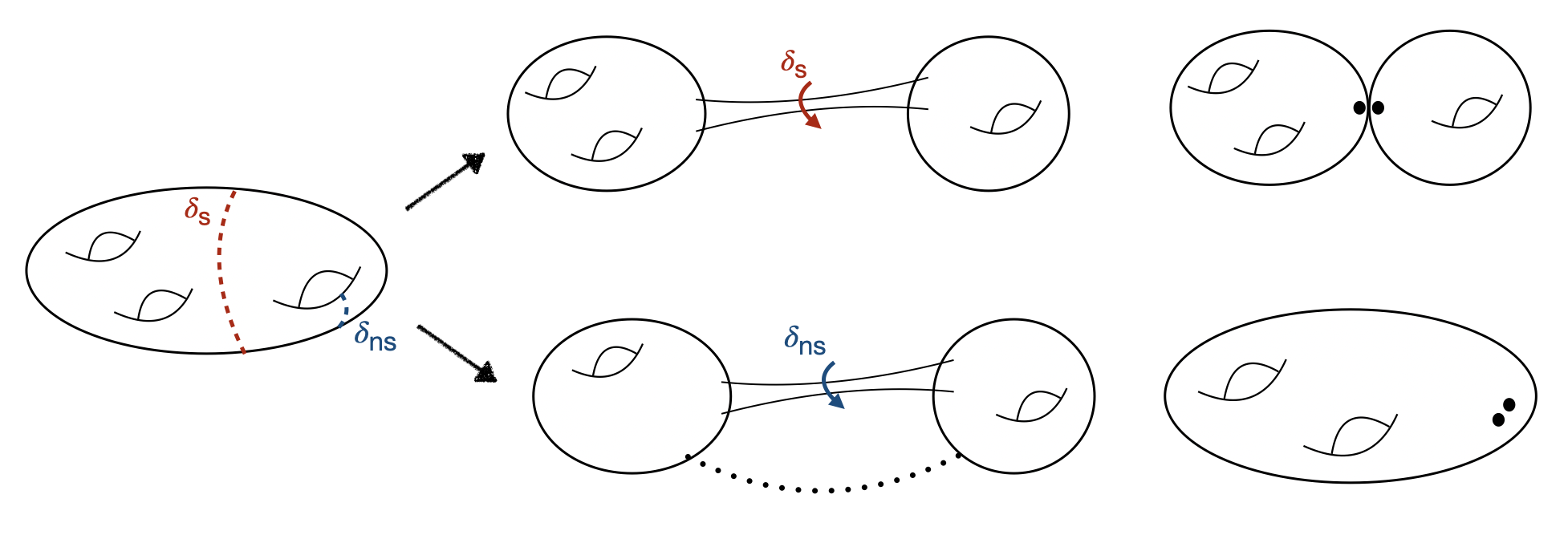}
    \caption{Separating and non-separating degenerations.  The separating
    degeneration has trivial Picard--Lefschetz monodromy on compact homology but
    may be detected by relative periods and open paths.  The non-separating
    degeneration is already detected by compact Picard--Lefschetz monodromy.}
    \label{fig:sepnonsep-node-unified}
\end{figure}

Let \(\overline X\) be a smooth marked curve with puncture divisor
\(D=\{p_1,\ldots,p_n\}\), and let
\[
X=\overline X\setminus D
\]
be the corresponding open curve. The standard description of nodal
degenerations of curves
\cite{HarrisMorrison:1998,Gaiotto:2009we} distinguishes two topological
types: separating and non-separating; see
Figure~\ref{fig:sepnonsep-node-unified}.

\paragraph{Separating nodes.}
\label{subsec:separating-nodes-unified}

A separating boundary divisor corresponds to a nodal degeneration in which the
curve splits into two connected components.  Equivalently, one pinches a
separating simple closed curve on the marked curve.  Schematically, the
limiting curve has the form
\be
    X_{g,n}
    \longrightarrow
    X_{g_L,n_L+1}\cup X_{g_R,n_R+1},
    \qquad
    g_L+g_R=g,
    \qquad
    n_L+n_R=n .
\ee
Here the original marked points are distributed between the two components,
and the additional marked point on each component is the preimage of the node
in the normalization.  The nodal curve is recovered by identifying these two
new marked points.  In compact homology, the vanishing cycle of a separating
degeneration is homologically trivial\footnote{Roughly, a cycle is homologically trivial if it is the boundary of a higher-dimensional surface.}.

\paragraph{Non-separating nodes.}

A non-separating boundary divisor corresponds to a nodal degeneration in which the curve remains connected, but its genus is reduced by one. Equivalently, one pinches a non-separating simple closed curve, thereby degenerating one handle. The normalization of the limiting curve is schematically
\be
    X_{g,n}
    \longrightarrow
    X_{g-1,n+2}.
\ee
The two additional marked points are the two preimages of the node in the
normalization, and the nodal curve is recovered by identifying them.  In this
case, the vanishing cycle is a nontrivial compact homology cycle before the
degeneration.

The relative-cohomology description
\cite{peters2008mixed} incorporates the cycles associated with marked-point
collisions and the motion of marked points. It fits into the exact sequence
\be
   0
   \longrightarrow
   H^0(D;\mathbb Z)/H^0(\overline X;\mathbb Z)
   \longrightarrow
   H^1(\overline X,D;\mathbb Z)
   \longrightarrow
   H^1(\overline X;\mathbb Z)
   \longrightarrow
   0 .
   \label{eq:relative-cohomology-exact-general}
\ee
Dually, relative homology contains both closed cycles on \(\overline X\) and
open paths whose endpoints lie on the marked points.  If the curve is genus \(g\) with
\(n\) marked points, then the compact part has rank
\be
    \operatorname{rank}H^1(\overline X;\mathbb Z)=2g,
\ee
while
\be
   H^0(D;\mathbb Z)/H^0(\overline X;\mathbb Z)
    \cong
    \mathbb Z^{n-1}
\ee
accounts for the \(n-1\) independent relative directions.  Therefore
\be
    \operatorname{rank}H^1(\overline{X},D;\mathbb Z)=2g+n-1 .
\ee
Thus \(H^1(\overline{X},D;\mathbb{Z})\) contains the ordinary compact periods of the curve together
with open periods measuring relative positions of punctures. 

\paragraph{Separating and non-separating coordinates.} To describe the degeneration and its monodromy explicitly, we now
recall the coordinates associated with these two types of nodes. For the analytic theory of compact Riemann surfaces used below, we follow \cite{fay2006theta}. Choose a symplectic basis of compact cycles
$ A_1,\ldots,A_g,\;B_1,\ldots,B_g$,
with
\be
    A_i\cdot A_j=0,
    \qquad
    B_i\cdot B_j=0,
    \qquad
    A_i\cdot B_j=\delta_{ij}.
\ee
If \(\omega_i\), \(i=1,\ldots,g\), are normalized holomorphic one-forms, then
the compact periods are
\be
    \Omega_{ij}=\int_{B_j}\omega_i .
\ee
In addition, an open path $\Gamma_{ab}:p_b\longrightarrow p_a$, with
$\partial\Gamma_{ab}=p_a-p_b$,
defines \emph{Abel--Jacobi coordinates}
\be
    u^i_{ab}=\int_{\Gamma_{ab}}\omega_i ,
\ee
which is a useful way to encode the
relative positions of marked points when \(g>0\).

 This distinction is important in degenerating families.  A \emph{non-separating}
degeneration is detected by compact \(H^1\): the vanishing cycle is a nontrivial
compact cycle, and the corresponding degeneration can be seen in the compact
period matrix, schematically
\be
    q_{ns}\sim e^{2\pi i\Omega} .
\ee

By contrast, a \emph{separating} vanishing cycle is homologically trivial in compact
\(H_1(\overline X;\mathbb Z)\).    In a local collision description, the separating coordinate can be
described using the prime form \cite{fay2006theta},
\be
    q_s\sim E(p,q\mid\Omega),
\ee
where
\be
    E(p,q\mid\Omega)
    =
    \frac{
    \theta[\Delta]\!\left(u_{pq}\mid\Omega\right)
    }
    {h_\Delta(p)h_\Delta(q)} .
\ee
Here \(\theta[\Delta](u\mid\Omega)\) is the Riemann theta function with theta
characteristic \(\Delta\), and \(h_\Delta(p)\), \(h_\Delta(q)\) are local
half-differentials associated with the same theta characteristic; see Appendix
\ref{app:theta-prime-form}. The above expression is a local coordinate representative. Near the collision locus, \(u^i_{pq}\propto
z(p)-z(q)\), and
\be
    E(p,q\mid\Omega)\sim z(p)-z(q).
\ee
This is the sense in which the marked-point collision degeneration considered
here is naturally captured by relative, rather than purely compact, period data.

\subsubsection{Base monodromy as a Dehn twist}\label{base mon}
Fix a base point \(q_0\in \Delta^\ast\).  A small loop around the boundary point is
\be
    \ell:\qquad q_0\longmapsto e^{2\pi i}q_0 .
\ee
This loop starts and ends at the same point of the base, but it acts
nontrivially on the fiber \(X_{q_0}\).  To see its action, write $q=|q|e^{i\varphi}$ and
    $u=re^{i\theta_u}$.
Using \eqref{eq:generic-plumbing-t} yields
\be
    v=\frac{q}{u}
    =
    \frac{|q|}{r}e^{i(\varphi-\theta_u)}.
\ee
Hence the angular coordinate on the \(v\)-side is related to the angular
coordinate on the \(u\)-side by
$
\theta_v=\varphi-\theta_u$.
When one goes once around the cusp,
\be
q\longmapsto e^{2\pi i}q,
\qquad
\varphi\longmapsto \varphi+2\pi ,
\ee
the two boundary circles of the plumbing annulus are reglued with one additional
full relative twist. We denote the core circle of the tube by \(\delta\).  Along the loop
\(\ell\), the absolute value \(|q|\) is fixed.  Therefore the length of the
annulus,
\be
    L\sim -\log|q_0|,
\ee
is unchanged.  What changes is the relative gluing angle between the two
branches of the node.  This change is precisely the local representative of a
\emph{Dehn twist} about the core circle \(\delta\).

\subsubsection{Picard--Lefschetz transformation} 
The Abelian projection of this monodromy is the Picard--Lefschetz
transformation.  For a compact
closed cycle \(\gamma\) one has locally \cite{PicardSimart:1897,Lefschetz:1924}
\be
    T_{\geom}(\gamma)
    =
    \gamma+(\delta\cdot\gamma)\delta .
    \label{eq:generic-picard-lefschetz}
\ee
The logarithm of the geometric monodromy is
\be
    N_{\geom}=\log T_{\geom} .
\ee
If the node is \emph{non-separating}, then \(\delta\) is nontrivial in compact homology
and \(N_{\geom}\neq0\) on \(H_1(\overline X_{q_0};\mathbb Z)\).  Choosing a dual
cycle \(\gamma\) with \(\delta\cdot\gamma=1\), one obtains
\be
    \delta\longmapsto\delta,
    \qquad
    \gamma\longmapsto\gamma+\delta .
\ee
In the basis \((\delta,\gamma)\), this is represented by
\be
    T_{\geom}
    =
    \begin{pmatrix}
        1&1\\
        0&1
    \end{pmatrix},
    \qquad
    N_{\geom}
    =
    \begin{pmatrix}
        0&1\\
        0&0
    \end{pmatrix},
    \qquad
    N_{\geom}^2=0 .
\ee
If the node is \emph{separating}, then \(\delta\) is trivial in compact homology, so
\be
    T_{\geom}=1,
    \qquad
    N_{\geom}=0.
\ee
Its plumbing dependence remains encoded in the coordinate \(q\); for
marked-point collision degenerations, it may also be detected in the relative
geometry of the pointed curve.

\subsubsection{Local non-Abelian monodromy}

We now associate a local \(G_{\mathbb C}\)-valued non-Abelian Hodge
monodromy with the geometric Dehn twist by computing the holonomy of the
fiberwise Hitchin--Simpson connection along the corresponding tube loop. On the smooth fibers
\(
    X_q=\pi^{-1}(q),\,
    q\in \Delta^\ast,
\)
we denote the corresponding \(G_{\mathbb C}\)-connections by \(\nabla_q\).
Near a boundary divisor, we use the
\emph{regular-singular} local model. After extending the bundle across the
relevant divisor, the connection has at most a logarithmic pole
\cite{deligne2006equations}. Near the tube, the
\(u\)-side of the plumbing annulus
\(
|q|/\epsilon<|u|<\epsilon
\)
approaches the punctured disk \(0<|u|<\epsilon\) as \(q\to0\). On the
normalization of the nodal limit, the corresponding marked point is \(u=0\).
We therefore write the leading part of the connection on the \(u\)-branch as
\be
    \nabla_q
    \simeq
    d-R\,\frac{du}{u},
    \label{eq:generic-tube-connection-u}
\ee
where \(R\in\mathfrak g_{\mathbb C}\) is the chosen logarithmic exponent of the
regular-singular asymptotic model.  After fixing a logarithmic branch, we regard its conjugacy class as \emph{fixed}
additional non-Abelian Hodge boundary data associated with the chosen family of
flat connections; it is not determined by the complex-structure degeneration
alone.

On a fixed smooth fiber \(X_q\), the parameter \(q\) is held fixed.  Therefore
\(dq=0\) along the fiber.  Since \(uv=q\), one obtains
${du}/{u}
    =
    -{dv}/{v}$.
%    \label{eq:generic-fixed-fiber-differentials}
Thus the same connection can also be written on the \(v\)-branch as
\be
    \nabla_q
    \simeq
    d+R\,\frac{dv}{v}.
    \label{eq:generic-tube-connection-v}
\ee
This is the same fiberwise connection, now written in the coordinate adapted to
the other branch of the normalization of the node.

On each fixed smooth fiber \(X_q\), the flat connection defines a monodromy
representation
\be
    \rho_q:\pi_1(X_q)\longrightarrow G_{\mathbb C}.
\ee
Let \(\delta\subset X_q\) denote the core loop of the plumbing annulus, for
example the circle \(|u|=\operatorname{const}\).  If \(\nabla_q=d+A_q\), then
\be
    M(\delta)=\rho_q(\delta)
    =P\exp\!\left(-\oint_{\delta}A_q\right).
\ee
In the logarithmic model \eqref{eq:generic-tube-connection-u}, this gives
\be
    M(\delta)=\exp(2\pi iR).
\ee

\begin{figure}[t]
\centering

\definecolor{twistred}{RGB}{162,31,29}
\definecolor{twistblue}{RGB}{20,35,130}

\tikzset{
    mainarrow/.style={->, thick, >=Stealth}
}

\begin{tikzpicture}[scale=0.95]

    \begin{scope}[xshift=2.5cm,yshift=1.5cm]

        \fill[gray!8] (0,1) -- (4,1) -- (4,-1) -- (0,-1) -- cycle;

        \draw[black!60] (0,1) -- (4,1);
        \draw[black!60] (0,-1) -- (4,-1);

        \draw[twistblue, line width=0.9pt]
            (0,0) ellipse [x radius=0.38, y radius=1.0];
        \draw[twistred, line width=0.9pt]
            (4,0) ellipse [x radius=0.38, y radius=1.0];

        \node[twistblue] at (-0.65,0) {\(u\)};
        \node[twistred] at (4.65,0) {\(v\)};

        \draw[dashed, thick] (2,0) ellipse [x radius=0.32, y radius=1.0];
        \node at (2.42,0.95) {\(\delta\)};

        \draw[gray!70, line width=1pt, -{Stealth[length=2mm,width=1.6mm]}]
            (0.08,0) -- (3.92,0);
        \node[gray!70] at (2,0.22) {\(\eta\)};

        \draw[
            twistred,
            line width=1.2pt,
            -{Stealth[length=2.2mm,width=1.7mm]},
            smooth,
            domain=0.08:3.92,
            samples=160
        ]
            plot (\x,{sin(360*(\x-0.08)/3.84)});
        \node[twistred] at (2,-1.35) {\(D_{\delta}(\eta)\simeq \delta\circ\eta\)};

    \end{scope}

    \begin{scope}[xshift=4.5cm,yshift=-4.1cm]
        \draw[black!60] (0,0) circle (1.55);

        \filldraw[black] (0,0) circle (1.5pt);
        \node at (0.30,0.28) {\(q=0\)};

        \filldraw[black] (1.15,0) circle (1.5pt);
        \node at (1.15,-0.32) {\(q_0\)};

        \draw[mainarrow] (1.15,0) arc[start angle=0,end angle=320,radius=1.15];
        \node at (0,1.95) {\(q\mapsto e^{2\pi i}q\)};
    \end{scope}

    \draw[dashed, -{Latex[length=2.5mm,width=1.8mm]}]
        (4.5,-0.5) -- (4.5,-1.5);
    \node at (4.75,-1.25) {\(\pi\)};

\end{tikzpicture}

\caption{A loop around the boundary point \(q=0\) in the base disk acts on the
nearby smooth curve as a Dehn twist of the plumbing cylinder.  The Dehn twist
inserts the core loop \(\delta\) into a path \(\eta\) crossing the tube.
Evaluating the flat connection on the inserted loop gives the local
non-Abelian Hodge monodromy.}
\label{fig:twistsingle}
\end{figure}
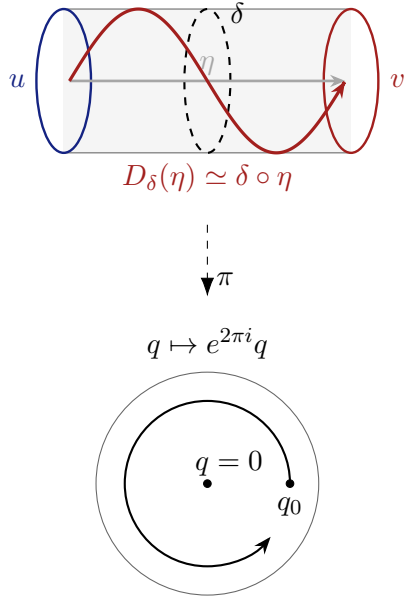

We evaluate the action of the Dehn twist on a path \(\eta\) crossing the
plumbing annulus. The twist inserts one copy of the core loop \(\delta\):
\be
    D_\delta(\eta)\simeq \delta\circ \eta .
\ee
We obtain the corresponding local non-Abelian Hodge contribution to the base
monodromy by evaluating the fiberwise flat connection on the inserted loop
\be
    M_{\rm NAH}(\delta)
    :=
    \rho_{q}(\delta)
    =
    \exp(2\pi iR).
\ee
This does not define an independent global representation of
\(\pi_1(B)\) into \(G_{\mathbb C}\). Rather, the base loop first acts
geometrically as the Dehn twist of the degenerating tube, and the local
\(G_{\mathbb C}\)-valued contribution is obtained by evaluating the UV-curve
flat connection on the corresponding tube loop (see figure \ref{fig:twistsingle}).

\paragraph{Dehn-twist action on non-Abelian holonomies.}
Having identified the local tube holonomy, we now make explicit how the
Dehn twist acts on flat holonomies. For the monodromy
representation \(\rho_q\),
\ba
    \rho_q(\delta)
    &\longmapsto&
    \rho_q(D_\delta(\delta))
    =
    \rho_q(\delta),
    \\
    \rho_q(\eta)
    &\longmapsto&
    \rho_q(D_\delta(\eta))
    =
   \rho_q(\delta) \rho_q(\eta).
\ea
Defining
\[
    M(\eta):=\rho_q(\eta),
    \qquad
    M(\delta):=\rho_q(\delta)=\exp(2\pi iR),
\]
this becomes
\be
    M(\eta)
    \longmapsto
  M(\delta)  M(\eta) 
    =
   \exp(2\pi iR) M(\eta).
    \label{eq:nonAbelian-picard-lefschetz}
\ee
This gives the non-Abelian monodromy action associated with the
Picard--Lefschetz transformation.  Its
Abelianization reduces to the usual insertion of the vanishing cycle,
\[
    \eta\longmapsto \eta+(\delta\cdot\eta)\delta .
\]
In the non-Abelian setting, this addition is replaced by \emph{ordered multiplication}
by the tube holonomy.  The order of the product depends on the convention for
composing paths and must be fixed together with the orientation convention for
the Dehn twist.  In this sense, base monodromy relabels the non-Abelian
holonomies associated with paths crossing the tube by multiplication with the
holonomy around the shrinking cycle, while the local tube holonomy itself is
unchanged.

\subsubsection{Transfer matrix and the asymptotic growth}

\begin{figure}
\begin{center}
\begin{tikzpicture}[x=1cm,y=1cm]

\draw[ubranch]
  (0.35,5.0) .. controls (0.7,4.1) and (0.95,4.0) .. (1.35,4.1)
  .. controls (1.45,4.12) and (1.52,4.12) .. (1.7,4.12);

\draw[vbranch]
  (5.65,5.0) .. controls (5.3,4.1) and (5.05,4.0) .. (4.65,4.1)
  .. controls (4.55,4.12) and (4.48,4.12) .. (4.3,4.12);

\draw[ubranch]
  (0.35,1.9) .. controls (0.7,2.8) and (0.95,2.9) .. (1.35,2.8)
  .. controls (1.45,2.78) and (1.52,2.78) .. (1.7,2.78);

\draw[vbranch]
  (5.65,1.9) .. controls (5.3,2.8) and (5.05,2.9) .. (4.65,2.8)
  .. controls (4.55,2.78) and (4.48,2.78) .. (4.3,2.78);

\shade[left color=lightublue,right color=lightvred]
  (1.7,2.78) rectangle (4.3,4.12);

\draw[hepblue]   (1.7,3.45) ellipse (0.28 and 0.67);
\draw[branchred] (4.3,3.45) ellipse (0.28 and 0.67);
\draw (1.7,4.12) -- (4.3,4.12);
\draw (1.7,2.78) -- (4.3,2.78);

\draw[mainarrow] (2.0,3.45) -- (4.0,3.45);

\node[font=\small] at (3.0,2.1) {$s_v=q^R s_u$};

\node[hepblue, font=\small] at (1.0,3.45) {$u$};
\node[branchred, font=\small] at (5.0,3.45) {$v$};

\end{tikzpicture}
\end{center}
\caption{Transfer of a flat section through the plumbing tube from the $u$-branch to the $v$-branch.}
\label{fig:transfer}
\end{figure}
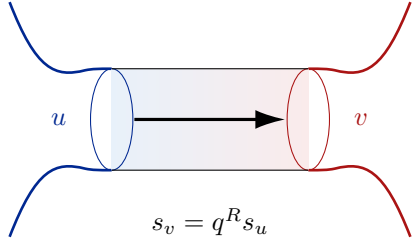

We next describe the leading parallel transport through the degenerating tube
in terms of a transfer matrix. In the local logarithmic tube model described
above, let \(s^{(q)}\) be a flat section of the vector bundle
\(V_q\rightarrow X_q\), with
\be
    \nabla_q s^{(q)}=0 .
\ee
Flat sections are the natural probes of the local monodromy because monodromy
is a property of parallel transport.  A flat section is locally constant with
respect to the connection, and after parallel transport around a loop one has
\be
    s^{(q)}\longmapsto M\,s^{(q)} .
\ee
Solving the leading flat-section equation gives
\(s^{(q)}(u)=u^R s_0\).
The \(u\)- and \(v\)-ends lie at \(u=\epsilon\) and
\(u=q/\epsilon\), respectively. Absorbing the fixed cutoff
factors into the normalization of the endpoint sections, we obtain
\be
    s_v^{(q)}
    =
    \mathcal T(q)\,s_u^{(q)},
    \qquad
    \mathcal T(q)=q^R .
    \label{eq:generic-tube-transfer}
\ee
The matrix \(\mathcal T(q)\) is the \emph{leading transfer matrix} across the long
plumbing tube; see Figure~\ref{fig:transfer}. It captures the leading singular
contribution to parallel transport in the local logarithmic model.

Finally, choosing a logarithm \(\alpha\) of \(M_s\), we decompose
\be
    R=\alpha+e,
    \qquad
    [\alpha,e]=0,
\ee
and obtain
\be
\begin{aligned}
    \mathcal T(q)
    &=
    q^R
    =
    q^\alpha
    \sum_{m=0}^{\ell-1}\frac{e^m}{m!}(\log q)^m,
    \qquad
    e^\ell=0 .
\end{aligned}
    \label{eq:generic-tube-growth-jordan}
\ee
The semisimple logarithmic exponent \(\alpha\) controls the power-law behavior in the plumbing parameter, while the nilpotent part \(e\) controls logarithmic growth. Since \(\exp(2\pi i\alpha)=M_s\), the monodromy determines \(\alpha\) only up to integral shifts of its eigenvalues. A chosen logarithmic exponent fixes the corresponding growth.  

We can also recover the tube holonomy associated with conformal-manifold
monodromy directly from the transfer matrix. Let \(\eta\) denote the segment
of a closed path crossing the tube from the \(u\)-branch to the \(v\)-branch,
and let \(\mathcal T(q)\) denote parallel transport along this segment.
Using \eqref{eq:generic-tube-transfer},
the Dehn twist replaces the crossing path, up to homotopy, by one with an
additional tube loop inserted. Since parallel transport along concatenated
paths is represented by the product of the corresponding transport matrices,
we obtain
\be
    \mathcal T(q)
    \longmapsto
    \rho_{q_0}(\delta)\,\mathcal T(q)
    =
    \exp(2\pi iR) \mathcal T(q).
\ee
This agrees with analytic continuation around the boundary point,
\be
   \mathcal T(e^{2\pi i}q)
    =
    (e^{2\pi i}q)^R
    =
    e^{2\pi iR}\mathcal T(q).
\ee
Thus the Dehn twist of the plumbing annulus acts on the tube-transfer matrix
by left multiplication with the local holonomy around the core circle
\(\delta\).

\subsection{Multi-parameter degeneration}
\label{subsec:multi-variable-nah-degeneration}

We now extend the one-parameter base-monodromy analysis to a boundary point
at which several degeneration parameters vanish simultaneously.  Let
\(
    I=\{1,\ldots,r\}
\)
be the set of boundary directions. In local coordinates
\((q_1,\ldots,q_d)\), the boundary divisors are \(D_i=\{q_i=0\}\), with
\(i\in I\).  The punctured local base is
\be
    B^{\ast}
    \simeq
    (\Delta^\ast)^r\times \Delta^{d-r},
    \qquad
    q_i\in \Delta^\ast ,\label{asy base}
\ee
where \(\Delta\) is a small complex disk and
\(\Delta^\ast=\Delta\setminus\{0\}\).
We consider a family
\be
    \pi:\mathcal X\longrightarrow B,
    \qquad
    X_q=\pi^{-1}(q),\label{map}
\ee
whose degenerations are locally described by plumbing equations
\be
    u_i v_i=q_i,
    \qquad
    i\in I .\label{plumbing}
\ee
For \(0<|q_i|\ll1\), the \(i\)-th plumbing region is a long annulus with core
loop \(\delta_i\subset X_q\). A loop around the \(i\)-th boundary divisor
\(D_i\) in the base is locally
\be
    \ell_i:\qquad q_i\longmapsto e^{2\pi i}q_i .
\ee
The loops \(\ell_i\) commute in the local base, and their induced actions on
the fibers are represented by Dehn twists about the core curves \(\delta_i\).
For each local base loop \(\ell_i\), we distinguish two related but distinct
monodromy transformations: \emph{geometric monodromy} acting on Abelian homology and
\emph{non-Abelian Hodge monodromy} obtained by evaluating the flat connection
around the corresponding tube loop.
\be
    \ell_i
    \longmapsto
    \left(
        T_{\geom}^{(i)},
        M_{\nah}^{(i)}
    \right).
\ee
The first, \(T_{\geom}^{(i)}\), is the geometric monodromy of the degenerating
curve.  Its logarithm \(N_{\geom}^{(i)}\) acts on Abelian homology and period
data, and the collection of such logarithms generates the geometric nilpotent
cone \({\rm Cone}_{\geom}(I)\).  The second, \(M_{\nah}^{(i)}\), is the local non-Abelian Hodge
monodromy obtained by evaluating the flat connection around the corresponding
tube loop \(\delta_i\), which gives \(\rho_q(\delta_i)\).  We stress that this
assignment is local; it does not define a full representation
\(
    \pi_1(B)\longrightarrow G_{\mathbb C}.
\)
Rather, a loop in the base first determines the corresponding
mapping-class-group action on the reference fiber, and the matrix
\(M_{\nah}^{(i)}\) is obtained only after evaluating the chosen fiberwise
monodromy representation on the tube loop \(\delta_i\) inserted by that
action.

\subsubsection{Geometric monodromy}
A loop around \(D_i\) returns to the same reference fiber but may act
nontrivially on its homology cycles. The resulting Abelian, or geometric,
monodromy is therefore
\be
    T_{\geom}^{(i)}
    \in
    \operatorname{Aut}\!\left(H_1(X_q;\mathbb Z)\right).
\ee
For a non-separating vanishing cycle \(\delta_i\), the
Picard--Lefschetz formula gives
\be
    T_{\geom}^{(i)}(\gamma)
    =
    \gamma+\left(\delta_i\cdot\gamma\right)\delta_i .
\ee
This local monodromy is unipotent, so its logarithm defines the nilpotent
generator
\be
    N_{\geom}^{(i)}=\log T_{\geom}^{(i)},
\ee
with
\be
    N_{\geom}^{(i)}(\gamma)
    =
    \left(\delta_i\cdot\gamma\right)\delta_i\,,
    \qquad
    \left(N_{\geom}^{(i)}\right)^2=0.
\ee
For a separating vanishing cycle,
\be
    N_{\geom}^{(i)}=0.
\ee
In this case, compact weight-one monodromy does not detect the degeneration.
Its plumbing dependence must be tracked separately and, for marked-point
collision divisors, may be described by relative or open-period data.

\paragraph{Abelian nilpotent cone.} The Picard--Lefschetz logarithms generate
the geometric nilpotent cone \cite{Kaplan1986}:
\be
    {\rm Cone}_{\geom}(I)
    =
    \left\{
    \sum_{i\in I} a_i N_{\geom}^{(i)}:
    a_i\geq0
    \right\} .\label{cone}
\ee
For a stable degeneration, the vanishing cycles can be chosen disjoint,
\be
    \delta_i\cdot\delta_j=0 .
\ee
Consequently the Picard--Lefschetz generators satisfy
\be
    N_{\geom}^{(i)}N_{\geom}^{(j)}=0.
\ee
Along an approach direction \(a=(a_i)_{i\in I}\), with \(a_i\geq0\), the
effective compact geometric logarithm is
\be
    N_{\geom}(a)=\sum_{i\in I}a_iN_{\geom}^{(i)},
    \qquad
   \left( N_{\geom}(a)\right)^2=0 .
\ee
Thus a multi-variable degeneration can contain several independent
non-separating directions, but for weight-one compact cohomology the geometric
nilpotency index is still at most two. If the approach direction has \(a_i>0\) for every active divisor \(i\in I\),
then \(N_{\mathrm{geom}}(a)=0\) only if
\(N_{\mathrm{geom}}^{(i)}=0\) for every \(i\in I\). If some \(a_i=0\),
the corresponding divisor is not probed by that one-parameter approach.

Near the boundary, the compact period data have the asymptotic form
\cite{Schmid1973,Kaplan1986}
\be
    \Pi(q)
    \sim
    \exp\!\left(
        \frac{1}{2\pi i}
        \sum_{i\in I}
        (\log q_i)N_{\geom}^{(i)}
    \right)\Pi_0 .
    \label{eq:generic-abelian-nilpotent-orbit}
\ee
Here \(\Pi(q)\) collectively denotes the compact periods, while
\(\Pi_0\) denotes their boundary data. Using the vanishing products above, this reduces to
\be
    \Pi(q)
    \sim
    \Pi_0
    +
    \frac{1}{2\pi i}
    \sum_{i\in I}
    (\log q_i)N_{\geom}^{(i)}\Pi_0 .
\ee
Along a one-parameter approach \(q_i=z^{a_i}\), with
\(a_i\in\mathbb Z_{\geq0}\) and \(z\to0\), the
logarithmic term is
\(
    \frac{\log z}{2\pi i}N_{\geom}(a)\Pi_0
\).
Hence the nilpotent-orbit approximation grows at most linearly in
\(\lvert\log|z|\rvert\), and its logarithmic part is absent when
\(N_{\geom}(a)\Pi_0=0\). Separating directions do not appear through a
nontrivial compact \(N_{\geom}^{(i)}\). Their dependence must instead be
tracked through the plumbing coordinates and, for marked-point collision
divisors, may also be described by relative or open periods.

\subsubsection{Non-Abelian monodromy}

We next determine the non-Abelian Hodge monodromy induced by the Dehn twist.
As a mapping-class-group element, the twist acts locally on a path
\(\eta\) crossing the \(i\)-th plumbing tube by composing it with one copy of
the core loop \(\delta_i\).  We write this schematically as
\be
    D_{\delta_i}(\eta)\simeq \delta_i\circ \eta .
\ee
The corresponding \(G_{\mathbb C}\)-valued matrix is obtained by applying the
monodromy representation of the flat connection to the inserted loop
\(\delta_i\).

On the fixed smooth curve \(X_q\), the chosen Hitchin--Simpson flat connection
defines a Betti monodromy representation
\be
    \rho_{q}:\pi_1(X_{q})\longrightarrow G_{\mathbb C}.
\ee
Near the \(i\)-th plumbing tube, we use the regular-singular local model,
whose leading logarithmic form is
\be
    \nabla_i
    \simeq
    d-R_i\frac{du_i}{u_i},
    \qquad
    R_i\in \mathfrak g_{\mathbb C}.
\ee
Writing \(\nabla_i=d+A_i\), we evaluate the holonomy around the core loop of
the tube
\be
    \rho_{q}(\delta_i)
    =
    P\exp\left(-\oint_{\delta_i}A_i\right)
    =
    \exp(2\pi iR_i).
\ee
We therefore obtain the local non-Abelian Hodge contribution
associated with the base monodromy around the boundary divisor
\be
    M_{\rm NAH}^{(i)}
    :=
    \rho_{q}(\delta_i)
    =
    \exp(2\pi iR_i).
\ee
Thus \(M_{\rm NAH}^{(i)}\) is not a separate global representation of
\(\pi_1(B)\) into \(G_{\mathbb C}\).  Rather, the base loop first acts
geometrically as a Dehn twist of the degenerating fiber, and the
\(G_{\mathbb C}\)-valued matrix is obtained by evaluating the fiberwise flat
connection on the tube loop selected by that Dehn twist.

\paragraph{Transfer matrices.}

We now consider the tube-wise transfer matrix and the action of
conformal-manifold monodromy on it. Let
\(\eta\) be a path crossing the \(i\)-th tube from the \(u_i\)-branch to the
\(v_i\)-branch. At leading order, flat sections on the two sides of the tube
are related by
\be
    s_{v_i}= \mathcal T_i(q_i)s_{u_i},
    \qquad
    \mathcal T_i(q_i)=q_i^{R_i}.
\ee
The base loop around \(D_i\) inserts one additional copy of the core loop into
the path crossing the tube. Hence
\be
    \mathcal T_i(q_i)
    \longmapsto
    \rho_{q_0}(\delta_i)\,\mathcal T_i(q_i)
    =
    \exp(2\pi iR_i)\mathcal T_i(q_i).
\ee
Analytic continuation of the local transfer matrix gives the same result:
\be
    \mathcal T_i(e^{2\pi i}q_i)
    =
    (e^{2\pi i}q_i)^{R_i}
    =
    \exp(2\pi iR_i)\mathcal T_i(q_i).
\ee
Thus the base loop around \(D_i\) acts geometrically by the Dehn twist of the
plumbing annulus. On the transfer matrix, this action is left multiplication
by the holonomy around the core loop.

\paragraph{Non-commutativity.}
One should distinguish the commutativity of local divisor loops from the
non-commutativity of the evaluated non-Abelian holonomies. Near a
normal-crossing boundary point, the base looks locally like
\((\Delta^\ast)^r\), so the coordinate loops
\(q_i\mapsto e^{2\pi i}q_i\) commute. The corresponding geometric Dehn twists
commute when the associated vanishing cycles are disjoint. However, the Betti
data on the fiber are group-valued. For generic loops \(\gamma_1\) and
\(\gamma_2\) on a fiber,
\be
    M(\gamma_a):=\rho_q(\gamma_a),
    \qquad a=1,2,
\ee
one can have
\be
    M(\gamma_1)M(\gamma_2)
    \neq
    M(\gamma_2)M(\gamma_1).
\ee
Thus commuting loops in the degeneration base do not imply commuting tube
holonomies. The natural non-Abelian object is the representation variety, or
character stack, together with the mapping-class-group action on it.

\subsubsection{Ordered multi-tube transport}

For each active divisor \(D_i\), we choose a regular-singular model in the
corresponding plumbing neighborhood:
\be
    \nabla_i
    \simeq
    d-R_i\frac{du_i}{u_i},
    \qquad
    R_i=\alpha_i+e_i,
    \qquad
    [\alpha_i,e_i]=0,
\ee
where \(\alpha_i\) is semisimple and \(e_i\) is nilpotent. We perform this
Jordan decomposition separately for each tube.  The corresponding
non-Abelian tube holonomies need not commute, and therefore no commutation relation is
assumed between exponent data associated with different tubes. At leading order,
the single-tube transfer factor is
\be
    \mathcal T_i(q_i)
    =
    q_i^{R_i}
    =
    q_i^{\alpha_i}
    \sum_{m=0}^{d_i}\frac{e_i^m}{m!}(\log q_i)^m,
    \qquad
    e_i^{d_i+1}=0.
    \label{eq:single-tube-factor-multivariable}
\ee
Hence \(\alpha_i\) controls the power-law behavior associated with the
\(i\)-th plumbing parameter and \(e_i\) controls the finite logarithmic
polynomial generated on that tube.

Let \(\mathfrak p\) be a path on the Riemann surface that crosses the
plumbing tubes in the order \(i_1,\ldots,i_k\). We decompose it
schematically as
\be
    \mathfrak p
    =
    \gamma_k\circ\rho_{i_k}\circ\gamma_{k-1}
    \circ\cdots\circ
    \gamma_1\circ\rho_{i_1}\circ\gamma_0,
\ee
where \(\rho_{i_a}\) crosses the \(i_a\)-th tube and \(\gamma_a\) runs along
the smooth part of the surface between successive tubes. In particular,
\(\gamma_0\) connects the initial point of \(\mathfrak p\) to the first tube,
while \(\gamma_k\) connects the last tube to its endpoint. The factors \(q_{i_a}^{R_{i_a}}\) give the leading contributions to parallel
transport across the tubes, obtained by matching the flat solutions on their
two sides, while parallel transport along the smooth segments \(\gamma_a\) is
represented by the matrices \(C_a(q)\). The leading parallel-transport matrix
along \(\mathfrak p\) is therefore
\be
    \mathcal T_{\mathfrak p}(q)
    =
    C_k(q)q_{i_k}^{R_{i_k}}C_{k-1}(q)
    \cdots
    C_1(q)q_{i_1}^{R_{i_1}}C_0(q).
    \label{eq:ordered-multitube-transfer}
\ee

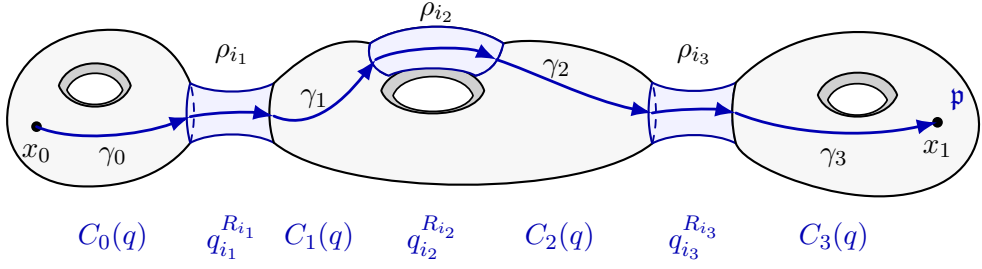
\begin{figure}[t]
    \centering
    \begin{tikzpicture}[
        x=0.82cm,
        y=0.82cm,
        >=Latex,
        line cap=round,
        line join=round,
        surface/.style={
            draw=black,
            fill=black!3,
            line width=0.8pt
        },
        hole rim/.style={
            draw=black,
            fill=black!18,
            line width=0.75pt
        },
        hole opening/.style={
            draw=black,
            fill=white,
            line width=0.65pt
        },
        tube/.style={
            draw=blue!60!black,
            fill=blue!5,
            line width=0.8pt
        },
        mouth/.style={
            draw=blue!60!black,
            dashed,
            line width=0.7pt
        },
        transport/.style={
            ->,
            draw=blue!70!black,
            line width=1.05pt
        }
    ]

    \path[surface]
        (0.05,-0.02)
        .. controls (0.10,0.83) and (0.84,1.36) ..
        (1.75,1.34)
        .. controls (2.47,1.33) and (2.83,0.91) ..
        (3.00,0.48)
        .. controls (3.05,0.20) and (3.05,-0.20) ..
        (3.00,-0.48)
        .. controls (2.64,-0.94) and (2.02,-1.22) ..
        (1.20,-1.18)
        .. controls (0.40,-1.14) and (0.00,-0.60) ..
        (0.05,-0.02)
        -- cycle;

    \path[hole rim]
        (0.88,0.44)
        .. controls (1.03,0.91) and (1.89,0.93) ..
        (2.04,0.44)
        .. controls (1.87,-0.01) and (1.06,-0.01) ..
        (0.88,0.44)
        -- cycle;

    \path[hole opening]
        (1.01,0.37)
        .. controls (1.14,0.72) and (1.78,0.74) ..
        (1.91,0.37)
        .. controls (1.77,0.06) and (1.15,0.06) ..
        (1.01,0.37)
        -- cycle;

    \path[tube]
        (3.00,0.48)
        .. controls (3.34,0.30) and (4.01,0.30) ..
        (4.35,0.48)
        .. controls (4.43,0.22) and (4.43,-0.22) ..
        (4.35,-0.48)
        .. controls (4.01,-0.30) and (3.34,-0.30) ..
        (3.00,-0.48)
        .. controls (2.92,-0.22) and (2.92,0.22) ..
        (3.00,0.48)
        -- cycle;

    \draw[mouth]
        (3.00,-0.48)
        .. controls (3.08,-0.20) and (3.08,0.20) ..
        (3.00,0.48);

    \draw[mouth]
        (4.35,-0.48)
        .. controls (4.27,-0.20) and (4.27,0.20) ..
        (4.35,0.48);

    \coordinate (tubeUL) at (5.90,1.12);
    \coordinate (tubeUR) at (8.05,1.12);
    \coordinate (tubeLR) at (7.60,0.60);
    \coordinate (tubeLL) at (6.20,0.60);

    \path[surface]
        (4.35,0.48)
        .. controls (4.92,1.03) and (5.42,1.24) ..
        (tubeUL)
        .. controls (6.43,1.48) and (7.52,1.48) ..
        (tubeUR)
        .. controls (8.55,1.25) and (9.90,1.08) ..
        (10.45,0.49)
        .. controls (10.55,0.20) and (10.55,-0.20) ..
        (10.45,-0.49)
        .. controls (9.35,-1.38) and (5.45,-1.38) ..
        (4.35,-0.48)
        .. controls (4.25,-0.20) and (4.25,0.20) ..
        (4.35,0.48)
        -- cycle;

    \path[hole rim]
        (6.08,0.36)
        .. controls (6.25,0.89) and (7.57,0.91) ..
        (7.74,0.36)
        .. controls (7.57,-0.13) and (6.25,-0.13) ..
        (6.08,0.36)
        -- cycle;

    \path[tube]
        (tubeUL)
        .. controls (6.43,1.48) and (7.52,1.48) ..
        (tubeUR)
        .. controls (8.00,0.91) and (7.90,0.68) ..
        (tubeLR)
        .. controls (7.34,0.76) and (6.48,0.76) ..
        (tubeLL)
        .. controls (5.92,0.70) and (5.84,0.91) ..
        (tubeUL)
        -- cycle;

    \path[hole opening]
        (6.27,0.28)
        .. controls (6.41,0.68) and (7.41,0.70) ..
        (7.56,0.28)
        .. controls (7.41,-0.06) and (6.41,-0.06) ..
        (6.27,0.28)
        -- cycle;

    \draw[mouth]
        (tubeUL)
        .. controls (5.84,0.91) and (5.92,0.70) ..
        (tubeLL);

    \draw[mouth]
        (tubeLR)
        .. controls (7.90,0.68) and (8.00,0.91) ..
        (tubeUR);

    \path[tube]
        (10.45,0.49)
        .. controls (10.79,0.31) and (11.46,0.31) ..
        (11.80,0.49)
        .. controls (11.88,0.22) and (11.88,-0.22) ..
        (11.80,-0.49)
        .. controls (11.46,-0.31) and (10.79,-0.31) ..
        (10.45,-0.49)
        .. controls (10.37,-0.22) and (10.37,0.22) ..
        (10.45,0.49)
        -- cycle;

    \draw[mouth]
        (10.45,-0.49)
        .. controls (10.53,-0.20) and (10.53,0.20) ..
        (10.45,0.49);

    \draw[mouth]
        (11.80,-0.49)
        .. controls (11.72,-0.20) and (11.72,0.20) ..
        (11.80,0.49);

    \path[surface]
        (11.80,0.49)
        .. controls (12.29,1.01) and (13.06,1.35) ..
        (13.94,1.30)
        .. controls (14.74,1.26) and (15.41,0.84) ..
        (15.65,0.19)
        .. controls (15.86,-0.38) and (15.58,-0.94) ..
        (15.00,-1.18)
        .. controls (14.26,-1.49) and (13.33,-1.31) ..
        (12.62,-1.04)
        .. controls (12.11,-0.84) and (11.85,-0.64) ..
        (11.80,-0.49)
        .. controls (11.72,-0.21) and (11.72,0.21) ..
        (11.80,0.49)
        -- cycle;

    \path[hole rim]
        (13.12,0.29)
        .. controls (13.29,0.76) and (14.22,0.78) ..
        (14.40,0.29)
        .. controls (14.22,-0.17) and (13.30,-0.18) ..
        (13.12,0.29)
        -- cycle;

    \path[hole opening]
        (13.28,0.21)
        .. controls (13.42,0.55) and (14.07,0.57) ..
        (14.24,0.21)
        .. controls (14.08,-0.11) and (13.43,-0.12) ..
        (13.28,0.21)
        -- cycle;

    \fill (0.52,-0.22) circle (2pt);
    \fill (15.05,-0.15) circle (2pt);

    \node[below=3pt] at (0.52,-0.22) {\(x_0\)};
    \node[below=3pt] at (15.05,-0.15) {\(x_1\)};

    \draw[transport]
        (0.52,-0.22)
        .. controls (1.12,-0.48) and (2.20,-0.36) ..
        (3.00,-0.06);

    \draw[transport]
        (3.00,-0.06)
        .. controls (3.38,0.02) and (3.98,0.02) ..
        (4.35,-0.05);

    \draw[transport]
        (4.35,-0.05)
        .. controls (4.88,-0.28) and (5.40,0.10) ..
        (5.99,0.86);

    \draw[transport]
        (5.99,0.86)
        .. controls (6.40,1.08) and (7.47,1.08) ..
        (7.94,0.86);

    \draw[transport]
        (7.94,0.86)
        .. controls (8.55,0.66) and (9.40,0.22) ..
        (10.45,0.00);

    \draw[transport]
        (10.45,0.00)
        .. controls (10.81,0.07) and (11.44,0.07) ..
        (11.80,0.00);

    \draw[transport]
        (11.80,0.00)
        .. controls (12.67,-0.35) and (13.89,-0.41) ..
        (15.05,-0.15);

    \node at (1.75,-0.70) {\(\gamma_0\)};
    \node at (5.00,0.23) {\(\gamma_1\)};
    \node at (8.91,0.72) {\(\gamma_2\)};
    \node at (13.38,-0.72) {\(\gamma_3\)};

    \node at (3.68,0.95) {\(\rho_{i_1}\)};
    \node at (6.98,1.62) {\(\rho_{i_2}\)};
    \node at (11.13,0.95) {\(\rho_{i_3}\)};

    \node[blue!70!black] at (1.75,-2.00)
        {\(C_0(q)\)};

    \node[blue!70!black] at (3.68,-2.00)
        {\(q_{i_1}^{R_{i_1}}\)};

    \node[blue!70!black] at (5.08,-2.00)
        {\(C_1(q)\)};

    \node[blue!70!black] at (6.91,-2.00)
        {\(q_{i_2}^{R_{i_2}}\)};

    \node[blue!70!black] at (8.96,-2.00)
        {\(C_2(q)\)};

    \node[blue!70!black] at (11.13,-2.00)
        {\(q_{i_3}^{R_{i_3}}\)};

    \node[blue!70!black] at (13.38,-2.00)
        {\(C_3(q)\)};

    \node[blue!70!black, above right=1pt]
        at (15.05,-0.15) {\(\mathfrak p\)};

    \end{tikzpicture}

    \caption{
        Parallel transport along a path crossing plumbing collars.
        The first and third collars correspond to separating degenerations,
        while the middle collar replaces the upper arm of a handle and
        represents a non-separating degeneration.
    }
    \label{fig:ordered-plumbing-transport}
\end{figure}

We assume that the off-tube matching matrices \(C_a(q)\) remain regular and
invertible as \(q\to0\). They can be absorbed into the neighboring tube
factors by repeatedly using
\be
    C\,q_i^{R_i}
    =
    \bigl(Cq_i^{R_i}C^{-1}\bigr)C
    =
    q_i^{C R_i C^{-1}}C .
\ee
Their information is therefore retained through conjugations of the exponent
matrices \(R_i\), while the remaining product of matching matrices contributes
only an overall regular factor. One may consequently write schematically
\be
    \mathcal T_{\mathfrak p}(q)
    \sim
    q_{i_k}^{R_{i_k}}\cdots q_{i_1}^{R_{i_1}}.
    \label{eq:ordered-multitube-singular}
\ee
This product is ordered, and its factors do not necessarily commute.
Expanding each factor using
\eqref{eq:single-tube-factor-multivariable} produces ordered mixed logarithmic
terms. Likewise, the nilpotent parts \(e_i\) need not commute. Consequently, unlike
in the commuting case, positive linear combinations of the \(e_i\) are not
guaranteed to be nilpotent, so there is no canonical non-Abelian nilpotent cone
associated with a generic multi-divisor cusp.

Although the plumbing base is multidimensional, the non-Abelian Hodge problem
is posed fiberwise on a one-complex-dimensional punctured curve. The disjoint
plumbing tubes carry distinct local exponents \(R_i\), and fiberwise flatness
by itself imposes no commutation relation among them. This contrasts with a
logarithmic flat connection on a higher-dimensional complex manifold with a
simple normal-crossing divisor, where integrability forces the residues of
intersecting divisor components to commute on their intersection
\cite{deligne2006equations}.

The Abelian nilpotent-orbit formula
\eqref{eq:generic-abelian-nilpotent-orbit} provides a useful point of
comparison with the non-Abelian tube asymptotics above. The operators
\(N_{\geom}^{(i)}\) control the geometric logarithmic growth, while \(e_i\)
controls the logarithmic part of each non-Abelian tube factor and \(\alpha_i\)
its semisimple power-law part. For several tubes, however, the non-Abelian
factors need not commute and therefore do not in general define a single
nilpotent cone. This comparison is closely related to the extension of
asymptotic Hodge-theoretic structures and growth formula to harmonic bundles developed by
Simpson and Mochizuki
\cite{Simpson1997MixedTwistor,Mochizuki2002,Mochizuki2003Asymptotic}.

\section{Application to type-\texorpdfstring{$A$}{A} class-\texorpdfstring{$\mathcal S$}{S} theories}
\label{sec:NAH for class S}

In this section we apply the local non-Abelian Hodge asymptotics developed in
Section~\ref{sec:nah-asymptotics} to the type-\(A\) class-\(\mathcal S\)
conformal-manifold cusps introduced in
Section~\ref{sec: class S material}. To each degenerating tube we associate a 
decoration specified by its local monodromy data and compute the corresponding monodromy centralizer. As we explain in Section~\ref{subsec:centralizers-cusp}, this centralizer is
the reductive symmetry preserving the local flat tube holonomy. The weakly coupled gauge
algebra, on the other hand, is fixed independently by the class-\(\mathcal S\) fixture and
gluing data. A particular class of monodromy labels for which the weak
gauge algebra lies in the corresponding centralizer is discussed
at the end of Section~\ref{sec:examples}. At a multi-divisor cusp,
the decoration and centralizer data are assigned tube by tube.

In particular, we construct the decorated cusp label
\(\mathfrak D_I\) introduced in \eqref{eq:decorated-cusp-label} and explain
its components. In type \(A\), nilpotent orbits are classified by their Jordan types,
equivalently by partitions of \(N\). Once the semisimple monodromy is fixed,
the Jordan type of the nilpotent part provides a discrete label for the
non-Abelian Hodge decoration and organizes the resulting decorated cusps into
discrete types.

\subsection{Local monodromy framework at class-\texorpdfstring{$\mathcal S$}{S} cusps}
\label{subsec:classS-conformal-manifold-boundary}

We now specialize the general degeneration analysis of Section
\ref{sec:nah-asymptotics} to  class-\(\mathcal S\)
theories.  In this specialization, the punctured Riemann
surface \(X\) is identified with the UV curve \(C\), while the degeneration
base \(B\) is identified with the conformal manifold \(\mathcal M\).  Thus,
the generic local base, family map, and plumbing equations of
\eqref{asy base}, \eqref{map}, and \eqref{plumbing} are understood with
\(X\to C\) and \(B\to\mathcal M\).  The UV curve is a punctured Riemann
surface
\be
    C_{g,n}=C_g\setminus D,
    \qquad
    D=\{z_1,\ldots,z_n\}.
\ee
The conformal manifold \(\mathcal M\equiv\mathcal M_{g,n}\) is the
complex-structure moduli space of the punctured UV curve.

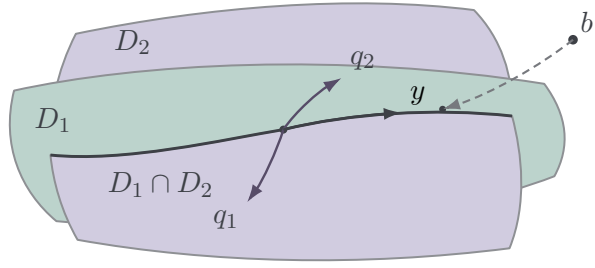
\begin{figure}[t]
    \centering
    \begin{tikzpicture}[
        x=1.12cm,
        y=1.12cm,
        >=Latex,
        sheet/.style={line width=0.75pt, line join=round},
        coordinate arrow/.style={-{Latex[length=2.4mm,width=1.6mm]},
                                 line width=0.9pt},
        limit arrow/.style={-{Latex[length=2.3mm,width=1.55mm]},
                            densely dashed, line width=0.8pt}
    ]
        \definecolor{modulijade}{RGB}{188,211,204}
        \definecolor{moduliviolet}{RGB}{209,204,224}
        \definecolor{moduliaccent}{RGB}{87,76,104}
        \definecolor{moduliink}{RGB}{61,64,70}

        \coordinate (L) at (-2.75,-0.18);
        \coordinate (P) at (0.00,0.12);
        \coordinate (R) at (2.70,0.28);

        \path[sheet, draw=moduliink!58, fill=moduliviolet]
            (L)
            .. controls (-1.88,-0.27) and (-0.72,0.00) .. (P)
            .. controls (0.78,0.31) and (1.78,0.38) .. (R)
            .. controls (2.72,0.76) and (2.57,1.20) .. (2.25,1.53)
            .. controls (0.82,1.71) and (-0.90,1.57) .. (-2.45,1.25)
            .. controls (-2.67,0.82) and (-2.78,0.34) .. cycle;

        \path[sheet, draw=moduliink!58, fill=modulijade]
            (-3.18,0.58)
            .. controls (-1.72,0.90) and (0.66,1.04) .. (3.08,0.66)
            .. controls (3.36,0.28) and (3.39,-0.08) .. (3.20,-0.43)
            .. controls (1.88,-0.94) and (-0.30,-1.18) .. (-2.68,-0.96)
            .. controls (-3.10,-0.61) and (-3.34,-0.06) .. cycle;

        \path[sheet, draw=moduliink!58, fill=moduliviolet]
            (L)
            .. controls (-1.88,-0.27) and (-0.72,0.00) .. (P)
            .. controls (0.78,0.31) and (1.78,0.38) .. (R)
            .. controls (2.82,-0.17) and (2.88,-0.76) .. (2.67,-1.28)
            .. controls (0.92,-1.50) and (-0.85,-1.46) .. (-2.43,-1.18)
            .. controls (-2.62,-0.85) and (-2.72,-0.49) .. cycle;

        \draw[moduliink, line width=1.10pt]
            (L)
            .. controls (-1.88,-0.27) and (-0.72,0.00) .. (P)
            .. controls (0.78,0.31) and (1.78,0.38) .. (R);

        \fill[moduliink] (P) circle (1.5pt);

        \draw[coordinate arrow, draw=moduliaccent]
            (P) .. controls (-0.08,-0.14) and (-0.25,-0.49) .. (-0.43,-0.74)
            node[below left=-1pt, text=moduliink] {$q_1$};
        \draw[coordinate arrow, draw=moduliaccent]
            (P) .. controls (0.15,0.34) and (0.42,0.57) .. (0.69,0.73)
            node[above right=-1pt, text=moduliink] {$q_2$};
        \draw[coordinate arrow, draw=moduliink]
            (P) .. controls (0.43,0.23) and (0.96,0.29) .. (1.40,0.32)
            node[above right=-1pt] {$y$};

        \node[text=moduliink] at (-2.72,0.23) {$D_1$};
        \node[text=moduliink] at (-1.78,1.15) {$D_2$};
        \node[below=3pt, text=moduliink] at (-1.45,-0.17)
            {$D_1\cap D_2$};

        \coordinate (B) at (3.42,1.18);
        \coordinate (Q) at (1.88,0.36);
        \fill[moduliink] (B) circle (1.8pt);
        \node[above right=-1pt, text=moduliink] at (B) {$b$};
        \fill[moduliink] (Q) circle (1.3pt);
        \draw[limit arrow, draw=moduliink!70]
            (B) .. controls (3.03,0.87) and (2.44,0.54) .. (Q);

    \end{tikzpicture}
    \caption{Local view near a two-fold boundary stratum of the compactified
    conformal manifold.  The divisors \(D_1\) and \(D_2\) intersect along the
    indicated curve, parametrized by \(y\).  At a point of this intersection,
    the \(q_1\) and \(q_2\) directions are transverse to \(D_1\) and \(D_2\),
    respectively.  The point \(b\) approaches the intersection as the
    transverse coordinates vanish.}
    \label{fig:multi-cusp-local-coordinates}
\end{figure}

\paragraph{Asymptotic monodromy.} Near a boundary point of the Deligne--Mumford compactification, several
plumbing parameters can vanish simultaneously.  Let
\(D_i=\{q_i=0\}\) denote the corresponding boundary divisors and
\(D_I=\bigcap_{i\in I}D_i\) their intersection.  Near \(D_I\), we choose
local coordinates
\(q=(q_i)_{i\in I}\) and \(y=(y_a)_{a\in\bar I}\) on the conformal manifold,
where \(\bar I\) labels the remaining local coordinates.  The coordinates
\(q_i\) are transverse to the boundary divisors, while the coordinates
\(y_a\) parametrize directions along the boundary stratum \(D_I\).  Thus,
the \(y_a\) remain finite as \(q_i\to0\).  We denote a point in this local
chart by
\be
    b=(q,y),
\ee
see Figure~\ref{fig:multi-cusp-local-coordinates}.  For \(q_i\neq0\) for all
\(i\in I\), the corresponding smooth UV curve is the fiber
\(
    C_b=\pi^{-1}(b).
\)

On each smooth fiber \(C_b\), the class-\(\mathcal S\) Hitchin system is
equipped with Higgs-bundle data
\(
    (V_b,\Phi_b)
\).
Under the appropriate stability and regularity assumptions, non-Abelian
Hodge theory associates these data with a Hitchin--Simpson flat connection
\ba
    \nabla_{h,b}
    =
    D_{h,b}
    +
    \Phi_b
    +
    \Phi^{\dagger}_{h,b}.
\ea
The flatness of \(\nabla_{h,b}\) gives a fiberwise monodromy representation
\(\rho_C:\pi_1(C_b)\to G_{\mathbb C}\), where the dependence on \(b\) is
understood.

For each \(i\in I\), let \(U_i\subset C_b\) be the disjoint plumbing annulus
associated with the \(i\)-th degenerating tube.  We analyze the local
asymptotic behavior of the chosen fiberwise Hitchin--Simpson flat connection
tube by tube, using the regular-singular logarithmic model of Section
\ref{sec:nah-asymptotics}
\be
    \left.\nabla_{h,b}\right|_{U_i}
    \simeq
    d-R_i\frac{du_i}{u_i},
    \qquad
    R_i\in\mathfrak{sl}(N,\mathbb C).
\ee
Here \(R_i\) is a chosen logarithmic exponent associated with the \(i\)-th tube.

As discussed in Section~\ref{base mon}, a loop around the boundary divisor
\(D_i\) is locally given by
\be
    \ell_i:\qquad
    q_i\longmapsto e^{2\pi i}q_i,
    \qquad
    q_j\ (j\neq i)\ \text{and }y\ \text{fixed}.
    \label{Mloopsec5}
\ee
This loop is the \emph{conformal-manifold monodromy} associated with the
corresponding singular direction.  It acts geometrically on the UV curve by
a Dehn twist around the core cycle \(\delta_i\) of the \(i\)-th tube. In weak-coupling coordinates, one writes
\be
    q_i=e^{2\pi i\tau_i},
    \qquad
    \tau_2^{(i)}=\operatorname{Im}\tau_i
    =
    -\frac{1}{2\pi}\log|q_i|.
\ee
Thus \(q_i\to0\) corresponds to
\(\tau_2^{(i)}\to\infty\), and the loop \eqref{Mloopsec5} corresponds to
\(\tau_i\mapsto\tau_i+1\).  With this normalization, the cusp monodromy is
the weak-coupling \emph{\(T\)-transformation}.

The induced action on compact homology is the Picard--Lefschetz monodromy
\(
T_{\geom}^{(i)}=\exp N_{\geom}^{(i)}
\).
For a non-separating node,
\(N_{\geom}^{(i)}\neq0\) and
\(\bigl(N_{\geom}^{(i)}\bigr)^2=0\).
For a separating node,
\(N_{\geom}^{(i)}=0\) on compact \(H_1\); the degeneration may nevertheless
be visible in relative or open-period data.  The geometric logarithms
\(N_{\geom}^{(i)}\) generate the geometric nilpotent cone \eqref{cone} and
control the logarithmic growth of the compact period vectors.

We obtain the corresponding local non-Abelian Hodge contribution by
evaluating the fiberwise flat connection on the tube loop \(\delta_i\)
\be
    M_{\nah}^{(i)}
    =
    \rho_C(\delta_i)
    =
    \exp(2\pi iR_i).
\ee
We decompose the exponent \(R_i\) independently for each tube as
\be
    R_i=\alpha_i+e_i,
    \qquad
    [\alpha_i,e_i]=0,
\ee
where \(\alpha_i\) is semisimple and \(e_i\) is nilpotent. The corresponding
single-tube transfer matrix is
\(
    \mathcal T_i(q_i)
    =
    q_i^{R_i}
\).
Its power-law growth is controlled by \(\alpha_i\), while its logarithmic
Jordan growth is controlled by \(e_i\). The exponent data associated with
distinct plumbing regions need not commute, and no commutation relation is
imposed between them.

For a path \(\mathfrak p\) on the UV curve crossing the plumbing tubes in the
order \(i_1,\ldots,i_k\), the corresponding multi-tube transport is given by
the ordered product \eqref{eq:ordered-multitube-transfer}.  Its singular
factors are \(q_i^{R_i}\), with
\(
    q_i^{R_i}=q_i^{\alpha_i}\exp(e_i\log q_i)
\).
Consequently, the ordered product controls the leading power-law and
logarithmic behavior of parallel transport through the collection of tubes.

Finally, we consider a segment \(\eta\) of a UV curve loop crossing the
\(i\)-th tube. With the path-composition convention
\(D_{\delta_i}(\eta)\simeq\delta_i\circ\eta\), and with
\(M(\delta_i):=M_{\nah}^{(i)}\), the Dehn twist acts on the corresponding
matrix as (see \eqref{eq:nonAbelian-picard-lefschetz})
\be
    M(\eta)
    \longmapsto
    \exp(2\pi iR_i)M(\eta).
    \label{eq:classS-nonAbelian-picard-lefschetz}
\ee
The Dehn twist preserves the conjugacy class of the local tube holonomy,
while changing the labeling of paths and the associated line-operator data
in the chosen weak-coupling description. In the class-\(\mathcal S\) duality
framework of \cite{Gaiotto:2009we}, more general mapping-class-group
transformations that change the pants decomposition relate distinct
S-dual weak-coupling frames.
 
\paragraph{Weakly coupled gauge group.}
The long plumbing tube has a direct gauge-theoretic interpretation.  Its
length satisfies
\be
    L_i\sim-\log|q_i|
    =2\pi\tau_2^{(i)},
    \qquad
    \tau_2^{(i)}\propto \frac{1}{g_i^2},
\ee
so that \(L_i\to\infty\) corresponds to \(g_i\to0\).  The geometry therefore
identifies a weak-coupling channel and determines how its coupling approaches
zero, but not which gauge algebra is weakly gauged through the tube.  We denote
this algebra by \(\mathfrak h_i\).  In a class-\(\mathcal S\) degeneration,
\(\mathfrak h_i\) is fixed independently by the puncture, fixture, and gluing
data.

The two monodromies provide complementary information. The geometric monodromy records the topological degeneration
of the UV curve, while the local non-Abelian Hodge monodromy records
the holonomy of the chosen flat connection around the tube core.
We introduce the corresponding reductive monodromy centralizer
in the following subsection.

\subsection{Monodromy centralizers}
\label{subsec:centralizers-cusp}

The continuous symmetry preserved by the tube holonomy
is described by its monodromy centralizer, the algebra of
infinitesimal transformations that commute with it.
We focus on its reductive part and denote the corresponding
compact real form by \(\mathfrak c_i\).

The local non-Abelian Hodge monodromy
\(M_i:=M_{\nah}^{(i)}\) is obtained by parallel transport of
\(\nabla_{h,b}\) around the circular direction of the tube and is defined
only up to conjugation. Under a change of local frame of the vector bundle by a
\(G_{\mathbb C}\)-valued function \(g\),
\be
    \nabla_{h,b}\longmapsto g\nabla_{h,b}g^{-1},
    \qquad
    M_i\longmapsto g(x_0)M_i g(x_0)^{-1},
\ee
where \(x_0\) is the chosen base point. Thus, after fixing a representative of
the monodromy conjugacy class, the transformations at the base point that
preserve this representative are precisely those in its centralizer. A change
of representative conjugates both the monodromy and its centralizer, so the
latter is well defined up to conjugation. 

The corresponding infinitesimal centralizer is
\be
    \mathfrak z_{i,\mathbb C}
    =
    \left\{
    X\in\mathfrak g_{\mathbb C}
    :
    M_i X M_i^{-1}=X
    \right\}.
    \label{eq:indexed-full-centralizer}
\ee
To describe it in terms of the asymptotic data, after choosing a branch of the
logarithm we write
\be
    M_i=\exp(2\pi iR_i),
    \qquad
    R_i=\alpha_i+e_i,
    \qquad
    [\alpha_i,e_i]=0,
\ee
where \(\alpha_i\) is semisimple and \(e_i\) is nilpotent. Let
\(
    M_{s,i}=\exp(2\pi i\alpha_i),
    \;
    M_{u,i}=\exp(2\pi i e_i)
\),
then
\(
    M_i=M_{s,i}M_{u,i}
\)
is the Jordan decomposition of \(M_i\). The simultaneous conjugacy class of \((M_{s,i},e_i)\), which is independent
of the logarithmic branch, defines the non-Abelian Hodge tube decoration,
whereas the chosen exponent \(\alpha_i\) specifies its power-law asymptotics.
The full centralizer of \(M_i\) is the common centralizer of \(M_{s,i}\) and
\(M_{u,i}\), or equivalently of \(M_{s,i}\) and \(e_i\), since
\(M_{u,i}=\exp(2\pi i e_i)\) with \(e_i\) nilpotent. 

The algebra \(\mathfrak z_{i,\mathbb C}\) need not be reductive. We denote
its reductive quotient by \footnote{When we regard \(\mathfrak c_{i,\mathbb C}\) as an actual subalgebra of
\(\mathfrak z_{i,\mathbb C}\), we choose a Levi factor, namely a reductive
subalgebra isomorphic to the reductive quotient
\(\mathfrak z_{i,\mathbb C}/\mathfrak r_u\).}
\be
    \mathfrak c_{i,\mathbb C}
    \simeq
    \bigl(\mathfrak z_{i,\mathbb C}\bigr)_{\rm red}.
    \label{eq:indexed-reductive-centralizer}
\ee
In the defining \(N\)-dimensional representation of \(SL(N,\mathbb C)\),
in which group elements act as \(N\times N\) matrices on \(\mathbb C^N\),
the semisimple monodromy decomposes the representation space into eigenspaces,
\[
    \mathbb C^N
    =
    \bigoplus_{\lambda} E_{i,\lambda},
    \qquad
    M_{s,i}|_{E_{i,\lambda}}
    =
    \lambda\,\mathrm{id}.
\]
Since \(M_{s,i}\) commutes with \(e_i\), each eigenspace
\(E_{i,\lambda}\) is preserved by \(e_i\), so
\(e_i|_{E_{i,\lambda}}\) may be put in Jordan form independently on each
eigenspace. Let
\(m_{i,\lambda,d}\) denote the number of size-\(d\) Jordan blocks of
\(e_i|_{E_{i,\lambda}}\).\footnote{A nilpotent Jordan block of size \(d\)
is denoted by \(J_d\).} As shown in \cite{FulmanGuralnick}, for type
\(A_{N-1}\) the reductive part of the centralizer takes the form
\be
    \mathfrak c_{i,\mathbb C}
    \simeq
    \mathfrak s
    \left(
        \bigoplus_{\lambda,d}
        \mathfrak{gl}\bigl(m_{i,\lambda,d},\mathbb C\bigr)
    \right).
    \label{eq:indexed-jordan-centralizer}
\ee
For fixed \((\lambda,d)\), the \(m_{i,\lambda,d}\) identical Jordan blocks
form a multiplicity space on which
\(\mathfrak{gl}(m_{i,\lambda,d},\mathbb C)\) acts by mixing the blocks while
preserving both \(M_{s,i}\) and \(e_i\). The additional transformations along
and between the Jordan chains belong to the non-reductive part of the
centralizer and are removed in passing to the reductive quotient.

The corresponding compact real form is
\be
    \mathfrak c_i
    \simeq
    \mathfrak s
    \left(
        \bigoplus_{\lambda,d}
        \mathfrak u\bigl(m_{i,\lambda,d}\bigr)
    \right).
    \label{eq:indexed-compact-centralizer}
\ee
Here \(\mathfrak s\) imposes the total tracelessness condition
\(
\sum_{\lambda,d}d\,\operatorname{tr}(X_{\lambda,d})=0
\).
Thus \(\mathfrak c_i\) is the compact reductive algebra associated with
continuous transformations preserving the tube monodromy. It belongs
to the non-Abelian Hodge decoration and neither represents the physical
puncture flavor algebra nor determines the weak gauging.

\paragraph{Multi-divisor boundary.}

For a conformal-manifold multi-divisor associated with a set \(I\) of
simultaneously degenerating but geometrically distinct tubes, we apply the
construction independently to each tube. We collect the corresponding
compact reductive algebras in
\be
    \mathfrak c_I
    =
    \bigoplus_{i\in I}\mathfrak c_i.
    \label{eq:multicusp-local-centralizer-product}
\ee

\paragraph{Example.}
Consider a one-divisor \(SL(5,\mathbb C)\) example with
\be
    \alpha
    =
    \mathrm{diag}(a,a,b,b,b),
    \qquad
    a-b\notin\mathbb Z,
    \qquad
    2a+3b=0 .
\ee
The defining representation decomposes as
\be
    \mathbb C^5
    =
    E_a\oplus E_b,
    \qquad
    E_a\simeq \mathbb C^2,
    \qquad
    E_b\simeq \mathbb C^3 .
\ee
Take the nilpotent part to be trivial on \(E_a\) and regular nilpotent on
\(E_b\),
\be
    e|_{E_a}=0,
    \qquad
    e|_{E_b}=J_3 .
\ee
Because \(a-b\notin\mathbb Z\), the spaces \(E_a\) and \(E_b\) are also
distinct eigenspaces of the semisimple monodromy. The Jordan type is
\([1,1]\) on \(E_a\) and \([3]\) on \(E_b\). Therefore
\be
    \mathfrak c_{\mathbb C}
    \simeq
    \mathfrak s
    \left(
        \mathfrak{gl}(2,\mathbb C)
        \oplus
        \mathfrak{gl}(1,\mathbb C)
    \right),
\ee
and the corresponding compact real form of the reductive centralizer is
\be
    \mathfrak c
    \simeq
    \mathfrak s
    \left(
        \mathfrak u(2)
        \oplus
        \mathfrak u(1)
    \right).
\ee
In words, the two trivial Jordan blocks in \(E_a\) can be rotated into one
another, while the single size-three Jordan block in \(E_b\) contributes only
an Abelian factor, subject to the overall tracelessness condition.

\subsection{The decorated cusp label}
\label{subsec:classification-cusp-labels}

In this section, we formulate the decorated cusp label in terms of the
geometric boundary data, the local non-Abelian monodromy, and the
independently specified weak gauging. We first present the full label,
then review the classification of nilpotent orbits by Jordan types and
use it to define a coarser discrete label for type-\(A\)
theories.

\subsubsection{Full decorated cusp label}

We organize the local data attached to a multi-divisor boundary point of a general class-$\mathcal{S}$ theory into the
full decorated cusp label
\be
    \mathfrak D_I
    =
    \left(
    \{\epsilon_i\}_{i\in I},
    \{M_{s,i},e_i\}_{i\in I},
    \{N_{\geom}^{(i)}\}_{i\in I},
    \{\mathfrak h_i\}_{i\in I}
    \right).
\ee
Here \(\epsilon_i\in\{s,ns\}\) records whether the \(i\)-th active divisor is
separating or non-separating. The non-Abelian label \((M_{s,i},e_i)\) records the
semisimple monodromy and the nilpotent logarithm of its unipotent part. A
chosen logarithm \(\alpha_i\), with
\(M_{s,i}=\exp(2\pi i\alpha_i)\), determines the power-law factor in the
tube-transfer matrix, while \(e_i\) determines its logarithmic growth. The compact geometric data \(N_{\geom}^{(i)}\) detect only the
non-separating part of the degeneration. Separating directions remain present
through their plumbing parameters and have trivial compact Picard--Lefschetz
logarithm; for marked-point collision divisors, they may also be tracked by
relative or open-period data. The algebra \(\mathfrak h_i\) is a separate physical label of the
\(i\)-th tube, determined independently by the class-\(\mathcal S\)
fixture and gluing data.

Only the node types and the geometric Picard--Lefschetz data are intrinsic to
the underlying boundary. The pairs \((M_{s,i},e_i)\), and hence the
associated tube-wise monodromy centralizers \(\mathfrak c_i\) and their direct
sum \(\mathfrak c_I=\bigoplus_{i\in I}\mathfrak c_i\), belong to the chosen
non-Abelian Hodge decoration. One may also specify an approach direction
\(a=(a_i)_{i\in I}\), together
with a path \(\mathfrak p\) fixing the tube-crossing order.  These are extrinsic approach data and
are not part of the decorated boundary label.

\subsubsection{Nilpotent orbits and Jordan types}\label{Jordan types}
As reviewed in \cite{collingwood1993nilpotent}, nilpotent conjugacy classes in
\(\mathfrak{sl}_N(\mathbb C)\) are classified by partitions of \(N\). For a fixed
semisimple monodromy \(M_{s,i}\), we apply this classification separately in
each eigenspace to the commuting nilpotent part \(e_i\), up to conjugation
preserving \(M_{s,i}\). Combining the eigenspace multiplicities of the
semisimple monodromy with these Jordan types, we organize the local monodromy
decorations into discrete types.

Every nilpotent element \(e_{N\times N}\in \mathfrak{sl}(N,\mathbb{C})\) admits a
decomposition into \textit{Jordan blocks} of size \(\nu_i\), such that
\be 
N=\sum_{i=1}^{r} \nu_i\,,
\qquad 
\nu_1 \geq \nu_2 \geq \cdots \geq \nu_r > 0 .
\ee
Then
\be 
e=J_{\nu_1}\oplus J_{\nu_2}\oplus \cdots \oplus J_{\nu_r}.
\ee
This nilpotent orbit is denoted by
\( 
e_{[\nu_1,\dots,\nu_r]} 
\).
The nilpotency index $\ell$ is determined by the size of the largest block:
\be 
e^{\nu_1}=0,
\qquad
e^{\nu_1-1}\neq0 .
\ee
Thus, we have
\(
\ell=\nu_1\), see Table \ref{tab:nilpotent-jordan-types}. The orbit \([N]\) has the maximal nilpotency index and is called the
principal orbit.

\begin{table}[H]
\centering
\renewcommand{\arraystretch}{1.35}
\begin{tabular}{c|c|c}
\hline
Partition \(\nu \vdash N\)
&
Jordan block form
&
Nilpotency index
\\
\hline
\([N]\)
&
\(J_N\)
&
\(N\)
\\

\([N-1,1]\)
&
\(J_{N-1}\oplus J_1\)
&
\(N-1\)
\\

\([N-2,2]\)
&
\(J_{N-2}\oplus J_2\)
&
\(N-2\)
\\

\([N-2,1,1]\)
&
\(J_{N-2}\oplus J_1\oplus J_1\)
&
\(N-2\)
\\

\([2,1^{N-2}]\)
&
\(J_2\oplus J_1\oplus \cdots \oplus J_1\)
&
\(2\)
\\

\([1^N]\)
&
\(J_1\oplus \cdots \oplus J_1=0\)
&
\(1\) 
\\
\hline
\end{tabular}
\caption{Selected nilpotent Jordan block types for partitions of \(N\).}
\label{tab:nilpotent-jordan-types}
\end{table}

\subsubsection{Discrete type-\texorpdfstring{$A$}{A} labels}\label{sec: discrete}

We now specialize the cusp label to type-A theories. The resulting discrete
type is a coarser label obtained by retaining only the eigenspace
multiplicities and the Jordan type on each eigenspace, rather than the
continuous eigenvalue data. For each \(i\in I\), let \(M_{s,i}\) and \(e_i\) denote, respectively, the
semisimple part and the nilpotent logarithm of the local non-Abelian
monodromy.  We decompose the defining representation according to the
distinct eigenvalues of \(M_{s,i}\),
\be
    \mathbb C^N
    =
    \bigoplus_{\lambda} E_{i,\lambda},
    \qquad
    M_{s,i}\big|_{E_{i,\lambda}}
    =
    \lambda\,\mathrm{id}_{E_{i,\lambda}},
    \qquad
    m_{i,\lambda}
    =
    \dim E_{i,\lambda}.
\ee
Since \(e_i\) commutes with \(M_{s,i}\), it preserves each eigenspace.  In
type \(A\), as introduced in Section~\ref{Jordan types}, the restriction
\(e_i|_{E_{i,\lambda}}\) is characterized up to conjugation by a partition
\be
    \nu_{i,\lambda}\vdash m_{i,\lambda}.
\ee
In the notation of that section, the corresponding nilpotent orbit is denoted
by \(e_{[\nu_{i,\lambda}]}\), where \(\nu_{i,\lambda}\) denotes the full
Jordan partition on \(E_{i,\lambda}\).  Thus, the discrete type of the
\(i\)-th tube is specified by
\be
    \mathsf T_i
    =
    \left\{
        \left(m_{i,\lambda},\nu_{i,\lambda}\right)
    \right\}_{\lambda}.
\ee
As we will see in the examples, the eigenspace multiplicities may be displayed as
\(2+1+1\), for example, and the corresponding Jordan types as
\([2]\,[1]\,[1]\).  The actual eigenvalues of \(M_{s,i}\), as well as the
choice of logarithmic lift \(\alpha_i\), are not part of the discrete type.

At a multi-divisor boundary point, we combine the geometric and physical
data with these tube types into the discrete cusp label \footnote{In the examples, we suppress the superscript \(A\), since the
restriction to type-\(A\) theories is understood.}
\be
    \mathfrak D_I^{A}
    =
    \left(
        \{\epsilon_i\}_{i\in I},
        \{N_{\geom}^{(i)}\}_{i\in I},
        \{\mathsf T_i\}_{i\in I},
        \{\mathfrak h_i\}_{i\in I}
    \right),
    \qquad
    \epsilon_i\in\{s,ns\}.\label{eq: coarse-tube-label}
\ee
The node types \(\epsilon_i\) and the geometric logarithms
\(N_{\geom}^{(i)}\) describe the underlying boundary, whereas
\(\mathsf T_i\) gives its discrete non-Abelian Hodge monodromy type decoration.   The corresponding tube-wise
centralizer depends only on these eigenspace multiplicities and Jordan data
and is therefore determined equivalently from \(\mathsf T_i\), as described in
Section~\ref{subsec:centralizers-cusp}.
A particular class of these labels, satisfying
\eqref{eq:monodromy-gauge-full-stabilizer} separately for each
\(i\in I\), is discussed in
Section~\ref{subsec:monodromy-gauge-class}.

\subsection{Physical interpretation and observables}
\label{subsec:physical-interpretation-cusp-label}

The decorated cusp label \(\mathfrak D_I\) organizes monodromy data
arising from the Hitchin-system description of class-\(\mathcal S\)
theories. For fixed global Higgs-bundle data in the setup considered
here, including the prescribed puncture structure, the non-Abelian
Hodge correspondence determines the associated flat connection up to
gauge equivalence, and hence its tube monodromies up to conjugation.
These data provide the geometric basis for our refinement of
weak-coupling cusps. We now ask how the components of the cusp label are reflected
in physical observables. We first examine the \(S^4\) partition
function and the Zamolodchikov metric to identify the dependence
of the leading cusp behavior on the weak gauge algebra, and then
discuss candidates for probing the tube monodromy data.

The physical metric on the conformal manifold is the Zamolodchikov metric,
defined by the two-point function of exactly marginal operators. As shown in
\cite{Gerchkovitz:2014gta}, for a four-dimensional $\mathcal N=2$ SCFT this
metric is K\"ahler and satisfies
\be
    Z_{S^4}=e^{K/12},
    \qquad
    g_{i\bar j}
    =12\,\partial_i\bar\partial_{\bar j}\log Z_{S^4},
    \label{eq:s4-zamolodchikov-relation}
\ee
Thus, the direct marginal-operator two-point function and the $S^4$
partition function contain the same information about the
Zamolodchikov metric.

Using supersymmetric localization \cite{Pestun:2007rz}, near a weak-coupling
divisor with $q=e^{2\pi i\tau}$ and
$\tau_2=\operatorname{Im}\tau\to\infty$, the localized gauge zero mode lies
in $\mathfrak t_{\mathfrak h}/W_{\mathfrak h}$. The corresponding integral has
the schematic form
\be
    Z_{S^4}
    \simeq
    \frac{1}{|W_{\mathfrak h}|}
    \int_{\mathfrak t_{\mathfrak h}}
    d^r a\,
    \Delta_{\mathfrak h}(a)^2
    e^{-2\pi\tau_2(a,a)}
    \mathcal Z_{\rm gl}(a;q,\bar q),
    \label{eq:schematic-localization-cusp}
\ee
where $\mathcal Z_{\rm gl}$ contains the contributions from the sectors
attached to the two ends of the tube, together with perturbative and instanton
effects. Assume that the leading gluing factor is regular
and nonzero at the weak-coupling saddle:
\be
    \mathcal Z_{\rm gl}(a;q,\bar q)
    =\mathcal Z_{\rm gl}(0;0,0)
    +\mathcal O(a)+\mathcal O(q,\bar q),
    \qquad
    \mathcal Z_{\rm gl}(0;0,0)\neq0.
\ee
Rescaling $a=\widetilde a/\sqrt{\tau_2}$ and using
$\dim\mathfrak h=r+|\Delta(\mathfrak h)|$ then gives
\be
    Z_{S^4}
    \sim
    C_0\,\tau_2^{-\dim\mathfrak h/2}.
    \label{eq:leading-partition-function-dimh}
\ee
Consequently, in the above normalization,
\be
    g_{\tau\bar\tau}
    \sim
    \frac{3\,\dim\mathfrak h}{2\tau_2^2},
    \qquad
    ds^2
    \sim
    \frac{3}{2}\,\dim\mathfrak h\,
    \frac{dq\,d\bar q}{|q|^2(\log|q|)^2}.
    \label{eq:expected-cusp-metric}
\ee
The metric therefore takes the universal Poincar\'e cusp form.
Its leading coefficient is fixed by \(\dim\mathfrak h\), where
\(\mathfrak h\) is determined independently by the fixture and
gluing data.

\paragraph{Monodromy-sensitive observables.}
A natural follow-up question is which physical observables could
probe the tube monodromy data \((M_s,e)\). Candidates to investigate include correlation functions
of local operators, supersymmetric line and surface-defect
observables, and subleading corrections to the Zamolodchikov metric.
The AGT correspondence offers a further framework for investigating
such sensitivity through the two-dimensional conformal blocks and
correlation functions associated with partition functions and
defect insertions.
Identifying such observables and establishing their dependence on
the tube monodromy lies beyond the scope of this work and is left
for future investigation.

\section{Examples}
\label{sec:examples}
In this section, we illustrate the construction with type-\(A\) examples,
using the discrete labels introduced in Section~\ref{sec: discrete}.
These labels record the eigenspace multiplicities of the semisimple
monodromy and the Jordan type of the nilpotent part within each eigenspace.
The examples should be understood as local tube data: whether a given monodromy
decoration is realized in a complete class-\(\mathcal S\) construction
depends on the choice of fixtures and the corresponding global
Hitchin-system data.

\subsection{The four-punctured sphere: \texorpdfstring{$SL(3)$}{SL(3)} tube types}
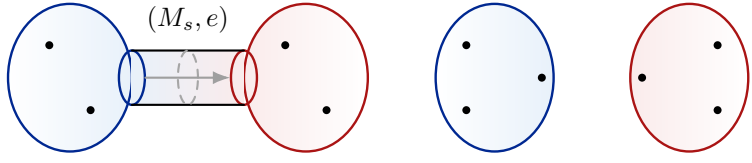
\begin{figure}[t]
\centering
\begin{tikzpicture}[
    x=0.78cm,
    y=0.78cm,
    line cap=round,
    line join=round,
    >=Latex
]

\shade[left color=white,right color=lightublue]
    (-5.0,0) ellipse (1.05 and 1.25);
\draw[ubranch,line width=0.9pt]
    (-5.0,0) ellipse (1.05 and 1.25);

\shade[left color=lightvred,right color=white]
    (-1.0,0) ellipse (1.05 and 1.25);
\draw[vbranch,line width=0.9pt]
    (-1.0,0) ellipse (1.05 and 1.25);

\foreach \x/\y in {
    -5.35/0.55,
    -4.65/-0.55,
    -1.35/0.55,
    -0.65/-0.55
}
    {\fill[black] (\x,\y) circle (1.5pt);}

\shade[left color=lightublue,right color=lightvred]
    (-3.95,-0.46) rectangle (-2.05,0.46);
\draw[hepblue,line width=0.9pt]
    (-3.95,0) ellipse (0.22 and 0.46);
\draw[branchred,line width=0.9pt]
    (-2.05,0) ellipse (0.22 and 0.46);
\draw[line width=0.8pt]
    (-3.95,0.46) -- (-2.05,0.46);
\draw[line width=0.8pt]
    (-3.95,-0.46) -- (-2.05,-0.46);

\draw[gray!70,dashed,line width=0.8pt]
    (-3.0,0) ellipse (0.17 and 0.46);
\draw[->,gray!75,line width=0.8pt]
    (-3.72,0) -- (-2.28,0);

\node[font=\small] at (-3.0,0.98) {$(M_s,e)$};

\shade[left color=white,right color=lightublue]
    (2.2,0) ellipse (1.0 and 1.25);
\draw[ubranch,line width=0.9pt]
    (2.2,0) ellipse (1.0 and 1.25);

\shade[left color=lightvred,right color=white]
    (5.5,0) ellipse (1.0 and 1.25);
\draw[vbranch,line width=0.9pt]
    (5.5,0) ellipse (1.0 and 1.25);

\foreach \x/\y in {1.72/0.55,1.72/-0.55,5.98/0.55,5.98/-0.55}
    {\fill[black] (\x,\y) circle (1.5pt);}

\fill[black] (3.0,0) circle (1.5pt);
\fill[black] (4.7,0) circle (1.5pt);

\end{tikzpicture}
\caption{Separating degeneration of \(C_{0,4}\): for \(0<|q|\ll1\), the left
and right sides are joined by a plumbing tube \(uv=q\), carrying decoration
\((M_s,e)\) and core cycle \(\delta\); at \(q=0\), they degenerate into left
and right \(C_{0,3}\) components meeting at a node.}
\label{fig:round-tube}
\end{figure}

For the four-punctured sphere, which we have  already introduced in Section~\ref{sec:Illustrative_example}, the conformal manifold is
\(
    \mathcal M_{0,4}
    \simeq
    \mathbb P^1_\lambda \setminus \{0,1,\infty\}.
\)
The three cusps are locally described by
\be
    q_0=\lambda,
    \qquad
    q_1=1-\lambda,
    \qquad
    q_\infty=\lambda^{-1}.
\ee
Each cusp corresponds to a separating degeneration of the UV curve.
Near a chosen cusp, we denote the local coordinate simply by \(q\),
so that the plumbing equation takes the form \(uv=q\).
For \(0<|q|\ll1\), the plumbing region forms a long tube connecting the two three-punctured spheres
(see Figure~\ref{fig:round-tube}),
\be
    C_{0,4}\longrightarrow C_{0,3}\cup C_{0,3}.
\ee

We use the regular-singular logarithmic tube model
\eqref{eq:generic-tube-connection-u}. The leading transfer matrix \eqref{eq:generic-tube-growth-jordan}
has power-law and logarithmic factors controlled by the semisimple
and nilpotent parts \(\alpha\) and \(e\) of the logarithmic exponent
\(R\), respectively. A loop around the cusp induces a Dehn twist
about the tube core \(\delta\), while the holonomy of the
Hitchin--Simpson connection around \(\delta\) supplies the local
non-Abelian monodromy data. The full decorated cusp label is
\be
    \mathfrak D_0
    =
    \left(
        \mathrm{sep};
        \ M_s,e;
        \ N_{\geom}=0;
        \ \mathfrak h
    \right),
\ee
where \((M_s,e)\) records the tube monodromy and \(\mathfrak h\) is
the weak gauge algebra fixed independently by the fixture and gluing
data. The compact Picard--Lefschetz monodromy is trivial because the
degeneration is separating.

We now specialize the complex structure group and its Lie algebra to
\be
    G_{\mathbb C}=SL(3,\mathbb C),
    \qquad
    \mathfrak g_{\mathbb C}=\mathfrak{sl}_3 .
\ee
The semisimple part of the local monodromy decomposes the fundamental
representation as
\be
    \mathbb C^3
    =
    \bigoplus_{\lambda}E_\lambda,
    \qquad
    M_s|_{E_\lambda}
    =
    \lambda\,\mathrm{id}_{E_\lambda}.
\ee
Since \(e\) commutes with \(M_s\), it preserves each eigenspace
\(E_\lambda\) of \(M_s\). The discrete local data are therefore specified by the dimensions \(\dim E_\lambda\) of the eigenspaces, together with the Jordan type
of the restriction \(e|_{E_\lambda}\) in each eigenspace. For \(SL(3)\), the
possible eigenspace decomposition patterns are
\(
    1+1+1,
    \,
    2+1,\text{and }
    3
\).

Table~\ref{tab:sl3-four-punctured-tube-types} lists the six resulting
coarse tube types entering the cusp label \eqref{eq: coarse-tube-label}.
It also gives the compact real forms of their reductive centralizers,
denoted by \(\mathfrak c(\mathsf T^{(3)}_a)\) and computed from the block
multiplicities using \eqref{eq:indexed-compact-centralizer}.
The last column records the maximal power of \(\log q\) generated by
\(\exp(e\log q)\).

\begin{table}[t]
\centering
\renewcommand{\arraystretch}{1.35}
\small
\begin{tabular}{c|c|c|c|c}
\hline
type
&
\(M_s\)-eigenspaces
&
Jordan type of \(e\)
&
\shortstack{monodromy centralizer\\
\(\mathfrak c(M_s,e)\)}
&
max.\ power of \(\log q\)
\\
\hline
\(\mathsf T^{(3)}_1\)
&
\(1+1+1\)
&
\([1]\,[1]\,[1]\)
&
\(\mathfrak u(1)^2\)
&
\(0\)
\\
\hline
\(\mathsf T^{(3)}_2\)
&
\(2+1\)
&
\([1,1]\,[1]\)
&
\(\mathfrak{su}(2)\oplus\mathfrak u(1)\)
&
\(0\)
\\
\(\mathsf T^{(3)}_3\)
&
\(2+1\)
&
\([2]\,[1]\)
&
\(\mathfrak u(1)\)
&
\(1\)
\\
\hline
\(\mathsf T^{(3)}_4\)
&
\(3\)
&
\([1,1,1]\)
&
\(\mathfrak{su}(3)\)
&
\(0\)
\\
\(\mathsf T^{(3)}_5\)
&
\(3\)
&
\([2,1]\)
&
\(\mathfrak u(1)\)
&
\(1\)
\\
\(\mathsf T^{(3)}_6\)
&
\(3\)
&
\([3]\)
&
\(0\)
&
\(2\)
\\
\hline
\end{tabular}
\caption{Local \(SL(3)\) non-Abelian Hodge tube types at a cusp of the
four-punctured sphere. The \(M_s\)-eigenspace column records the
multiplicities of the distinct eigenvalues of the semisimple monodromy, and
\(\mathfrak c(M_s,e)\) is the compact real form of the corresponding
reductive monodromy centralizer.}
\label{tab:sl3-four-punctured-tube-types}
\end{table}

The following representative choices of \((M_s,e)\) illustrate these types.
For \(\mathsf T^{(3)}_1\), we may take
\be
    M_s
    =
    \operatorname{diag}(\mu_1,\mu_2,\mu_3),
    \qquad
    \mu_1\mu_2\mu_3=1,
    \qquad
    e=0,
\ee
with \(\mu_1,\mu_2,\mu_3\) pairwise distinct. The reductive centralizer is
the diagonal torus in \(SL(3,\mathbb C)\), whose compact real form has Lie
algebra
\(
    \mathfrak c=\mathfrak u(1)^2
\).

For \(\mathsf T^{(3)}_2\) and \(\mathsf T^{(3)}_3\), we may take
\be
    M_s
    =
    \operatorname{diag}(\mu_1,\mu_1,\mu_2),
    \qquad
    \mu_1^2\mu_2=1,
    \qquad
    \mu_1\neq\mu_2.
\ee
The choices \(e=0\) and \(e=J_2\oplus J_1\) give
\(\mathsf T^{(3)}_2\) and \(\mathsf T^{(3)}_3\), respectively.
For \(e=0\), the nilpotent part is trivial on the two-dimensional
\(M_s\)-eigenspace, and
\(
    \mathfrak c
    =
    \mathfrak{su}(2)\oplus\mathfrak u(1)
\).
For \(e=J_2\oplus J_1\), the nontrivial Jordan block removes the
corresponding \(\mathfrak{su}(2)\) factor from the reductive centralizer,
leaving
\(
    \mathfrak c=\mathfrak u(1)
\).

For the remaining three types, \(M_s\) has a single three-dimensional
eigenspace, so
\be
    M_s=\lambda\,\mathrm{id},
    \qquad
    \lambda^3=1.
\ee
The choices
\(
    e=0
\),
\(
    e=J_2\oplus J_1
\), and
\(
    e=J_3
\)
give \(\mathsf T^{(3)}_4\), \(\mathsf T^{(3)}_5\), and
\(\mathsf T^{(3)}_6\), respectively. Their reductive centralizers are
\be
    \mathfrak c\!\left(\mathsf T^{(3)}_4\right)=\mathfrak{su}(3),
    \qquad
    \mathfrak c\!\left(\mathsf T^{(3)}_5\right)=\mathfrak u(1),
    \qquad
    \mathfrak c\!\left(\mathsf T^{(3)}_6\right)=0.
\ee
In particular, the principal nilpotent orbit \([3]\) has trivial reductive
centralizer.

These representatives must also admit a commuting semisimple logarithm
\(
    \alpha\in\mathfrak{sl}(3,\mathbb C)
\).
For \(\mathsf T^{(3)}_4\) and \(\mathsf T^{(3)}_5\), every
\(\lambda\) satisfying \(\lambda^3=1\) admits such a logarithm. For the
principal case \(\mathsf T^{(3)}_6\), however, a semisimple \(\alpha\)
commuting with \(J_3\) must vanish. Hence \(M_s=\mathrm{id}\), and therefore
\(\lambda=1\).

The table records only the eigenspace multiplicities of \(M_s\) and the
Jordan type of \(e\), not the actual eigenvalues of \(M_s\). Each row
therefore specifies a coarse discrete label of the form introduced in
Section~\ref{sec: discrete}.
For a cusp \(c\in\{0,1,\infty\}\) carrying local
tube type \(\mathsf T^{(3)}_a\), the corresponding
cusp label is
\be
    \mathfrak D_c\big(\mathsf T^{(3)}_a\big)
    =
    \left(
        \mathrm{sep};
        \ N_{\geom}^{(c)}=0;
        \ \mathsf T^{(3)}_a;
        \ \mathfrak h_c
    \right),
    \qquad
    a=1,\ldots,6.
\ee

\paragraph{Weak-coupling channels.}
In the \(SL(3)\) setting of
Table~\ref{tab:sl3-four-punctured-tube-types},
two inequivalent physical weak-coupling channels arise for the
\(A_2\) theory on a sphere with two full and two simple punctures. The full
and simple punctures have Nahm partitions \([1^3]\) and \([2,1]\), with
flavor symmetries \(SU(3)\) and \(U(1)\), respectively, while their Hitchin
partitions are \([3]\) and \([2,1]\). These puncture partitions characterize
the defects at the marked points and should be distinguished from the Jordan
type of \(e\), which specifies the nilpotent part of the tube monodromy.

In the ordinary SQCD degeneration channel, the four-punctured sphere splits
into two three-punctured spheres, each containing two full punctures and one
simple puncture. Gluing the two internal full punctures gauges a diagonal
\(SU(3)\). Each component contributes three fundamental hypermultiplets with
respect to the gauged group, giving \(SU(3)\) SQCD with six fundamentals.

In the other inequivalent degeneration channel, the two full punctures lie on
the same component. Together with the full puncture associated with the
degenerating tube, this component gives the \(T_3\) theory, namely the
interacting \(E_6\) SCFT. The second component contributes one fundamental
hypermultiplet of \(SU(2)\), while the tube corresponds to a weakly coupled
\(SU(2)\) vector multiplet gauging an \(SU(2)\) subgroup of the \(E_6\)
flavor symmetry. This is the Argyres--Seiberg duality frame
\cite{Argyres:2007cn,Gaiotto:2009we}. These channels illustrate the physical gauging entering
the cusp label.
Table~\ref{tab:sl3-four-punctured-tube-types}
lists the complete set of coarse local \(SL(3)\) monodromy
types, providing an algebraic starting point for a future
study of which types admit global realizations with the
prescribed puncture data.

\subsection{The two-punctured torus: \texorpdfstring{$SL(4)$}{SL(4)} tube types}

We now consider the two-punctured torus, which, unlike the four-punctured
sphere, admits both separating and non-separating degenerations. Its conformal manifold also has codimension-two boundary points
at which two tubes degenerate simultaneously.
We write the UV curve as
\be
    C_{1,2}
    =
    E_{\tau_E}\setminus\{z_1,z_2\},
    \qquad
    E_{\tau_E}
    =
    \mathbb C/(\mathbb Z+\tau_E\mathbb Z).
\ee
Here \(\tau_E\) is the complex-structure modulus of the elliptic curve, while
\(z_1\) and \(z_2\) are the marked points. The conformal manifold has complex dimension
\(
    \dim_{\mathbb C}\mathcal M_{1,2}=2.
\)

\begin{figure}[t]
    \centering
    \includegraphics[
        width=0.9\textwidth,
        clip
    ]{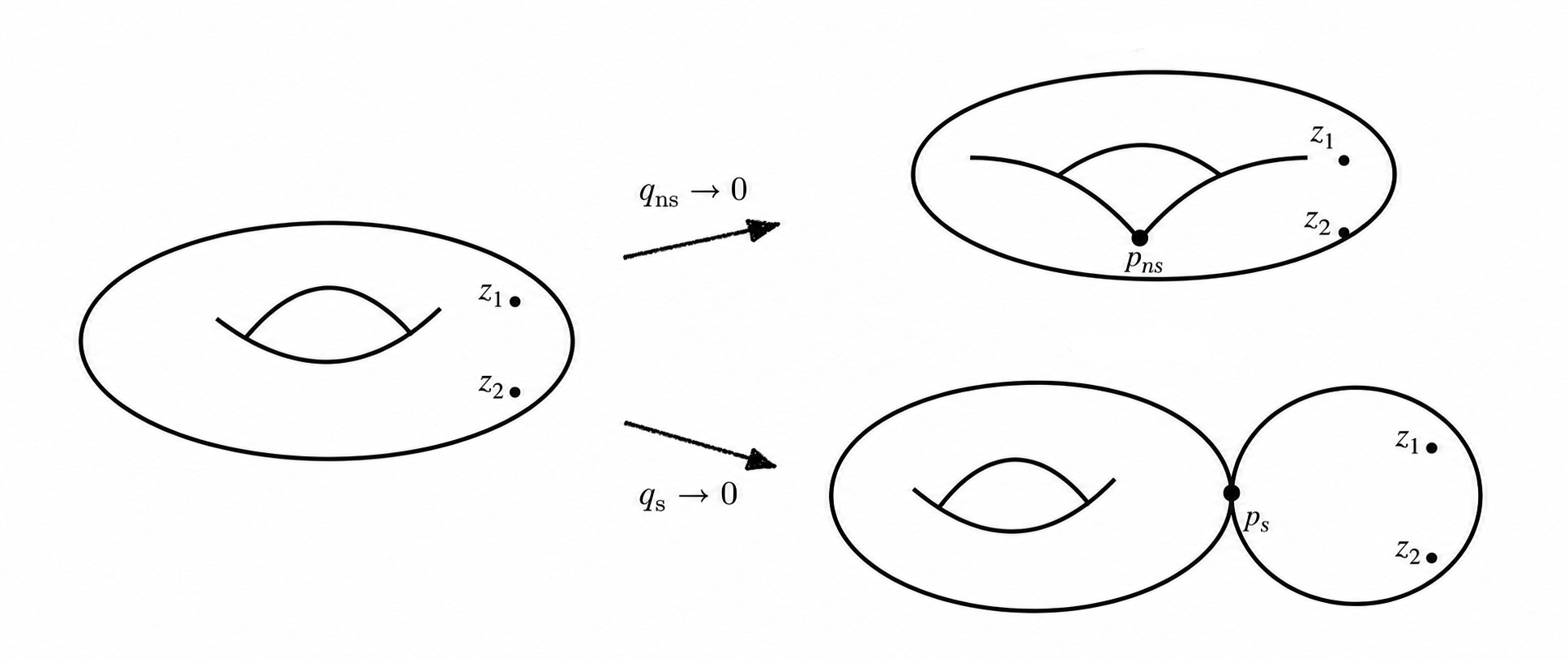}
    \caption{The two degeneration directions of the two-punctured torus.
    The upper branch is the non-separating degeneration
    \(\tau_E\to i\infty\), while the lower branch is the separating
    degeneration \(\Delta z=z_1-z_2\to0\).}
    \label{fig:two-punctured-torus-degenerations}
\end{figure}

As discussed in Section~\ref{subsec:one-variable-nah-degeneration}, the
period data are naturally organized by the relative cohomology group
\(
    H^1(E_{\tau_E},\{z_1,z_2\};\mathbb Z).
\)
They consist of the two compact torus periods and one independent relative
period between the marked points. Choosing the holomorphic one-form
\(\omega\) to be normalized on the \(A\)-cycle, we have
\be
    \int_A\omega=1,
    \qquad
    \int_B\omega=\tau_E,
    \qquad
    u_{12}=\int_{z_2}^{z_1}\omega .
\ee
The modulus \(\tau_E\) controls the non-separating degeneration, while the
relative period \(u_{12}\) records the relative position of the punctures and
locally detects their collision in the separating degeneration. Thus, the
two-punctured torus provides a simple setting in which compact-cycle
Picard--Lefschetz monodromy and relative-period data can be distinguished
explicitly. We denote the corresponding separating and non-separating
boundary coordinates by \(q_s\) and \(q_{ns}\), respectively. These two
degeneration types are illustrated schematically in
Figure~\ref{fig:two-punctured-torus-degenerations}.

We now make the monodromy and transfer-matrix formulas of
Section~\ref{sec:NAH for class S} explicit for the two-punctured torus,
with \(i=s,ns\) labeling the separating and non-separating
degenerations and \(q_s,q_{ns}\) their respective boundary coordinates.

\paragraph{Separating divisor.}

The separating boundary divisor is
\(
    D_s=\{q_s=0\}
\).
At this divisor the two marked points collide, and the UV curve degenerates as
\be
    C_{1,2}
    \longrightarrow
    C_{1,1}\cup C_{0,3}.
\ee
The vanishing cycle separates off a three-punctured sphere.  

A convenient local coordinate for this degeneration is provided by the
elliptic specialization of the prime form \cite{fay2006theta}. For an
elliptic curve,
\be
    q_s
    \sim
    E(z_1,z_2\mid\tau_E)
    =
    \frac{\vartheta_1(z_1-z_2\mid\tau_E)}
    {\vartheta_1'(0\mid\tau_E)} .
\ee
The elliptic identifications
\(z\sim z+m+n\tau_E\), with \(m,n\in\mathbb Z\), change the prime form
only by a nonvanishing factor. Hence its vanishing locus is independent
of the choice of representative for the puncture positions. Near the
collision locus we have
\(
    E(z_1,z_2\mid\tau_E)
    \sim
    z_1-z_2 \). Therefore \(q_s\to0\) describes the degeneration in which the
two punctures collide.

A loop in the conformal manifold encircling the separating divisor acts on the
local coordinate as
\(
   \ell_s: q_s\longmapsto e^{2\pi i}q_s 
\).
Geometrically, this corresponds to a Dehn twist about the separating
vanishing cycle. On the non-Abelian Hodge side, the flat connection carries
the corresponding tube holonomy
\be
    M_{\nah}^{s}
    =
    \exp(2\pi i R_s),
    \qquad
    R_s=\alpha_s+e_s,
    \qquad
    [\alpha_s,e_s]=0 .
\ee
The semisimple part \(\alpha_s\) controls the power-law behavior in \(q_s\),
while the nilpotent part \(e_s\) controls possible logarithmic growth. The
associated tube-transfer matrix is
\(
    \mathcal T_s(q_s)
    =
    q_s^{R_s}
\).

For the separating degeneration, the associated compact
Picard--Lefschetz monodromy is trivial
\(
    T_{\geom}^{s}=1,
    \,
    N_{\geom}^{s}=0
\).
Indeed, the separating vanishing cycle is homologically trivial in the
compact curve. This does not make the degeneration itself trivial: it remains
encoded by the plumbing coordinate \(q_s\), and the open-period data.

\paragraph{Non-separating divisor.}

The non-separating boundary divisor is
\(D_{ns}=\{q_{ns}=0\}\).
This degeneration pinches one of the compact cycles of the torus.  We choose
the \(A\)-cycle to be the vanishing cycle.  The normalization removes the
handle and produces a four-punctured sphere,
\be
    C_{1,2}
    \longrightarrow
    C_{0,4},
\ee
where two punctures come from the normalization of the node and two are the
original marked points.
The corresponding degeneration coordinate is the usual elliptic parameter
\be
    q_{ns}
    =
    e^{2\pi i\tau_E}.
\ee
Thus \(q_{ns}\to0\) is the large-complex-structure limit, \(
\operatorname{Im}\tau_E\longrightarrow\infty
\).

A loop in the conformal manifold encircling the non-separating divisor acts on the
local coordinate as
\(
  \ell_{ns}:  q_{ns}\longmapsto e^{2\pi i}q_{ns}
\),
or equivalently
\(
    \tau_E\longmapsto \tau_E+1 
\). Geometrically, this is the Dehn twist about the vanishing \(A\)-cycle.
On the non-Abelian Hodge side, the corresponding local tube holonomy is
obtained by evaluating the flat connection around this cycle,
\be
    M_{\nah}^{ns}
    =
    \exp(2\pi iR_{ns}),
    \qquad
    R_{ns}=\alpha_{ns}+e_{ns},
    \qquad
    [\alpha_{ns},e_{ns}]=0 .
\ee
The corresponding transfer matrix to leading order is
\(
    \mathcal T_{ns}(q_{ns})
    =
    q_{ns}^{R_{ns}} 
\).

In contrast with the separating case, the compact geometric monodromy is
nontrivial.  If \(A\) is the vanishing cycle and \(B\) is its dual cycle, then
the Picard--Lefschetz transformation is
\be
    A\longmapsto A,
    \qquad
    B\longmapsto B+A .
\ee
In the ordered basis \((A,B)\),  this gives the \(2\times2\) block
\be
    T_{\geom}^{ns}
    =
    \begin{pmatrix}
        1&1\\
        0&1
    \end{pmatrix}
    =
    1+N_{\geom}^{ns},
    \qquad
    N_{\geom}^{ns}
    =
    \begin{pmatrix}
        0&1\\
        0&0
    \end{pmatrix},
    \qquad
    (N_{\geom}^{ns})^2=0 .
\ee
For this genus-one example, the matrix \(T_{\geom}^{ns}\) is the full compact
Picard--Lefschetz block.

The Dehn twist action on non-Abelian holonomies can be seen explicitly in the \(A,B\) basis.  Let \(M_A\) and \(M_B\) be the flat holonomies around the \(A\)-
and \(B\)-cycles.  One has
\be
    M_A\longmapsto M_A,
    \qquad
    M_B\longmapsto  M_A M_B.
\ee
If \(
    M_A=\exp(2\pi iR_{ns}),\) then the \(B\)-cycle holonomy transforms as
\(
    M_B
    \mapsto
   \exp(2\pi iR_{ns}) M_B 
\).

\paragraph{The double cusp.}

We now approach the codimension-two boundary point
\be
    D_s\cap D_{ns}
    =
    \{q_s=0,\ q_{ns}=0\}.
\ee
Geometrically, the degeneration can be viewed as
\be
    C_{1,2}
    \xrightarrow{\ q_{ns}\to0\ }
    C_{0,4}
    \xrightarrow{\ q_s\to0\ }
    C_{0,3}\cup C_{0,3}.
\ee
The first step pinches the handle, producing a four-punctured sphere;
the second separates this sphere into two trinions. Reversing the order
changes only the intermediate degeneration: both orders reach the
same stable curve. In its normalization, the preimages of the
separating node lie on different trinions, while those of the
non-separating node lie on the same trinion. Thus, the double cusp
corresponds to a \emph{pair-of-pants decomposition} of the
two-punctured torus.

The two local base loops, \(\ell_s\) and \(\ell_{ns}\), commute,
but the corresponding tube holonomies need not. Each tube carries
the local logarithmic exponent and Jordan decomposition introduced
above; no commutation relation between \(R_s\) and \(R_{ns}\) is assumed.
For example, let \(\mathfrak p\) be a transport path that crosses first the
non-separating tube and then the separating tube. Its double-cusp transfer operator is
\be
    \mathcal T_{\mathfrak p}(q_s,q_{ns})
    =
    C_2(q)q_s^{R_s}C_1(q)q_{ns}^{R_{ns}}C_0(q).
    \label{eq:torus-ordered-double-transfer}
\ee
Expanding the nilpotent exponentials in the two tube factors gives
a finite sum involving \(q_s^{\alpha_s}\), \(q_{ns}^{\alpha_{ns}}\),
and powers of \(\log q_s\) and \(\log q_{ns}\), with the matrix factors
kept in their prescribed order. The compact geometric monodromy at
the double cusp is entirely controlled by the non-separating node:
\be
    N_{\geom}^{\rm tot}
    =
    N_{\geom}^{ns},
    \,
    (N_{\geom}^{\rm tot})^2=0. 
\ee
The separating divisor contributes nontrivially to the relative/open geometry
and to the non-Abelian tube data, but it contributes trivially to the compact
Picard--Lefschetz logarithm.

We now specialize to the type \(A_3\) theory,
\be
    G_{\mathbb C}=SL(4,\mathbb C),
    \qquad
    \mathfrak g_{\mathbb C}=\mathfrak{sl}_4 .
\ee
For a local tube, the semisimple monodromy \(M_s\) decomposes the defining
representation \(\mathbb C^4\) into its eigenspaces,
\be
    \mathbb C^4
    =
    \bigoplus_{\lambda}E_\lambda,
    \qquad
    M_s|_{E_\lambda}
    =
    \lambda\,\mathrm{id}_{E_\lambda}.
\ee
Since the nilpotent part \(e\) commutes with \(M_s\), it preserves each
eigenspace \(E_\lambda\). We therefore organize the local data by the
eigenspace decomposition of \(M_s\), together with the Jordan type of \(e\)
on each eigenspace. 

Table~\ref{tab:sl4-tube-types} lists the fourteen coarse
\(SL(4,\mathbb C)\) tube types entering the cusp label
\eqref{eq: coarse-tube-label}. Here we record the eigenspace
multiplicities of \(M_s\), rather than its actual eigenvalues.
The compact real forms of the reductive monodromy centralizers,
denoted by \(\mathfrak c(\mathsf T^{(4)}_a)\), are computed from the
block multiplicities \(m_{\lambda,d}\) using
\eqref{eq:indexed-compact-centralizer}.

\begin{table}[t]
\centering
\renewcommand{\arraystretch}{1.35}
\small
\begin{tabular}{c|c|c|c}
\hline
type
&
\(M_s\)-eigenspaces
&
\(e\)-type
&
\(\mathfrak c(M_s,e)\)
\\
\hline
\(\mathsf T^{(4)}_1\)
&
\(1+1+1+1\)
&
\([1]\,[1]\,[1]\,[1]\)
&
\(\mathfrak u(1)^3\)
\\
\hline
\(\mathsf T^{(4)}_2\)
&
\(2+1+1\)
&
\([1,1]\,[1]\,[1]\)
&
\(\mathfrak{su}(2)\oplus\mathfrak u(1)^2\)
\\
\(\mathsf T^{(4)}_3\)
&
\(2+1+1\)
&
\([2]\,[1]\,[1]\)
&
\(\mathfrak u(1)^2\)
\\
\hline
\(\mathsf T^{(4)}_4\)
&
\(2+2\)
&
\([1,1]\,[1,1]\)
&
\(\mathfrak{su}(2)_1\oplus\mathfrak{su}(2)_2\oplus\mathfrak u(1)\)
\\
\(\mathsf T^{(4)}_5\)
&
\(2+2\)
&
\([1,1]\,[2]\)
&
\(\mathfrak{su}(2)\oplus\mathfrak u(1)\)
\\
\(\mathsf T^{(4)}_6\)
&
\(2+2\)
&
\([2]\,[2]\)
&
\(\mathfrak u(1)\)
\\
\hline
\(\mathsf T^{(4)}_7\)
&
\(3+1\)
&
\([1,1,1]\,[1]\)
&
\(\mathfrak{su}(3)\oplus\mathfrak u(1)\)
\\
\(\mathsf T^{(4)}_8\)
&
\(3+1\)
&
\([2,1]\,[1]\)
&
\(\mathfrak u(1)^2\)
\\
\(\mathsf T^{(4)}_9\)
&
\(3+1\)
&
\([3]\,[1]\)
&
\(\mathfrak u(1)\)
\\
\hline
\(\mathsf T^{(4)}_{10}\)
&
\(4\)
&
\([1,1,1,1]\)
&
\(\mathfrak{su}(4)\)
\\
\(\mathsf T^{(4)}_{11}\)
&
\(4\)
&
\([2,1,1]\)
&
\(\mathfrak{su}(2)\oplus\mathfrak u(1)\)
\\
\(\mathsf T^{(4)}_{12}\)
&
\(4\)
&
\([2,2]\)
&
\(\mathfrak{su}(2)\)
\\
\(\mathsf T^{(4)}_{13}\)
&
\(4\)
&
\([3,1]\)
&
\(\mathfrak u(1)\)
\\
\(\mathsf T^{(4)}_{14}\)
&
\(4\)
&
\([4]\)
&
\(0\)
\\
\hline
\end{tabular}
\caption{Local \(SL(4)\) non-Abelian Hodge tube types. The
\(M_s\)-eigenspace column records the dimensions of the eigenspaces associated
with the distinct eigenvalues of the semisimple monodromy, while
\(\mathfrak c(M_s,e)\) denotes the compact real form of the corresponding
reductive monodromy centralizer.}
\label{tab:sl4-tube-types}
\end{table}

To specify the power-law factor \(q^\alpha\) in
\eqref{eq:generic-tube-growth-jordan}, we choose a semisimple logarithm
\(\alpha\in\mathfrak{sl}_4(\mathbb C)\) commuting with \(e\) and
satisfying \(M_s=\exp(2\pi i\alpha)\).
This choice is branch dependent and is not part of the coarse
discrete label. Requiring such a logarithm restricts the allowed
eigenvalues in three cases:
\be
    \mathsf T^{(4)}_6:\quad \lambda_1\lambda_2=1,
    \qquad
    \mathsf T^{(4)}_{12}:\quad \lambda^2=1,
    \qquad
    \mathsf T^{(4)}_{14}:\quad \lambda=1.
\ee
Here \(\lambda_1,\lambda_2\) are the two distinct eigenvalues in
\(\mathsf T^{(4)}_6\), while \(M_s=\lambda\,\mathrm{id}\) in the latter two
cases. These conditions refine the representatives of the corresponding
types without changing the discrete classification in
Table~\ref{tab:sl4-tube-types}.

\paragraph{Single-divisor cusp labels.}

The separating and non-separating divisors have the same possible local
non-Abelian Hodge tube types listed in Table~\ref{tab:sl4-tube-types}. They
differ by their compact geometric monodromy. For each row \(\mathsf T^{(4)}_a\), the
separating decorated cusp label is
\be
    \mathfrak D_s(\mathsf T^{(4)}_a)
    =
    \left(
    \mathrm{sep};
    \ N_{\geom}^{s}=0;
    \ \mathsf T^{(4)}_a;
    \ \mathfrak h_s
    \right),
    \qquad
    a=1,\ldots,14 .
\ee
The non-separating decorated cusp label is
\be
    \mathfrak D_{ns}(\mathsf T^{(4)}_a)
    =
    \left(
    \mathrm{nonsep};
    \ N_{\geom}^{ns}\neq0,
    \quad
    (N_{\geom}^{ns})^2=0;
    \ \mathsf T^{(4)}_a;
    \ \mathfrak h_{ns}
    \right),
    \qquad
    a=1,\ldots,14 .
\ee

\paragraph{Double-cusp labels.}

At the double cusp \(D_s\cap D_{ns}\), the chosen local monodromy data
consist of two types,
\be
    (M_{s,s},e_s)\in\mathsf T^{(4)}_a,
    \qquad
    (M_{s,ns},e_{ns})\in\mathsf T^{(4)}_b,
    \qquad
    a,b=1,\ldots,14.
\ee
No commutation relation between the corresponding logarithmic exponents
\(R_s\) and \(R_{ns}\) is imposed. These data
are accompanied by the compact geometric logarithm
\(
    N_{\geom}^{\rm tot}=N_{\geom}^{ns},
    \;
    (N_{\geom}^{\rm tot})^2=0 
\).

We label the local monodromy types at the double cusp by the ordered pair
\(
(\mathsf T^{(4)}_a,\mathsf T^{(4)}_b)
\),
where the first entry refers to the separating tube and the second
to the non-separating tube. The corresponding discrete cusp label is
\be
    \mathfrak D_{s,ns}
    =
    \big(
    (s,ns);
    \ N_{\geom}^{\rm tot}\neq0;
    \ \mathsf T^{(4)}_a,\mathsf T^{(4)}_b;
    \ \mathfrak h_s,\mathfrak h_{ns}
    \big).
\ee

\subsection{A particular class of monodromy labels}
\label{subsec:monodromy-gauge-class}

Among the general cusp labels constructed above,
it may be worthwhile to consider a particular class of
cusp labels in which the weak gauge algebra embeds in the
compact reductive monodromy centralizer. 
We select this class by imposing
\be
    \mathfrak h_i
    \subseteq
    \mathfrak c_i.
    \label{eq:monodromy-gauge-full-stabilizer}
\ee
A motivation for this class comes from flat adjoint sections
on the tube: a section with value \(X\) at a base point is
single-valued precisely when \(M_iXM_i^{-1}=X\).
With the Levi realization described in
Section~\ref{subsec:centralizers-cusp}, the inclusion places
the embedded generators of \(\mathfrak h_i\) in this invariant
algebra. The natural generalization to a multi-divisor cusp is to
impose
\eqref{eq:monodromy-gauge-full-stabilizer} separately on each tube, 
so that $\mathfrak h_I
    =
    \bigoplus_{i\in I}\mathfrak h_i
    \subseteq
    \bigoplus_{i\in I}\mathfrak c_i$.
The direct sum collects the separate inclusions associated with
the distinct tube neighborhoods.

Physical realization remains subject to the matter content,
flavor symmetries, and the four-dimensional conformality condition. Applied to the local types in
Table~\ref{tab:sl3-four-punctured-tube-types},
the inclusion retains only \(\mathsf T^{(3)}_4\) for
\(\mathfrak h=\mathfrak{su}(3)\), and
\(\mathsf T^{(3)}_2\) or \(\mathsf T^{(3)}_4\) for
\(\mathfrak h=\mathfrak{su}(2)\).
Determining which of these labels admit global
realizations with prescribed puncture and gluing data
is left to future investigation.

\section{Conclusions and outlook}\label{sec: conclusion}

In this paper, we refined the description of weak-coupling cusps
in type-\(A\) class-\(\mathcal S\) theories by combining the geometric
monodromy of the degenerating UV curve with local non-Abelian
monodromy data. The complex structure of the curve parametrizes
the exactly marginal couplings, while the associated Hitchin system
carries additional data encoding the Coulomb-branch geometry.
To incorporate local information from this structure, we used
non-Abelian Hodge theory to associate a Hitchin--Simpson flat
connection with the chosen Higgs bundle on the UV curve. We assigned the local monodromy of this flat connection to each
degenerating tube. 

To identify these monodromy data, we considered a loop
around the corresponding boundary divisor, which acts on the UV curve by a Dehn twist.  Its action on compact homology is described by
Picard--Lefschetz monodromy, which is trivial for a separating node
and has a nonzero nilpotent logarithm for a non-separating node.
The corresponding non-Abelian monodromy is assigned locally to
each tube as the holonomy of the fiberwise Hitchin--Simpson flat
connection around its core.

We then studied the asymptotic behavior of flat sections transported
through the degenerating tubes. Within the regular-singular
logarithmic tube model, the semisimple and nilpotent parts of the
chosen logarithmic exponent determine, respectively, the power-law
factor and the finite logarithmic polynomial in the leading
tube-transfer matrix. For simultaneous degenerations, we used a local logarithmic model
for each tube. Since the tube holonomies need not commute, we
expressed transfer through several tubes as an ordered product
of transfer matrices along the chosen path.
We also examined the leading Zamolodchikov metric near a
weak-coupling cusp. For a regular nonvanishing gluing factor,
its leading coefficient is fixed by the dimension of the
weak gauge algebra.

In type \(A\), semisimple eigenspace multiplicities and nilpotent
Jordan types organize the local monodromy data into coarse discrete
types and determine their reductive centralizers. We illustrated
this structure in the \(SL(3,\mathbb C)\) four-punctured-sphere and
\(SL(4,\mathbb C)\) two-punctured-torus examples, covering
separating, non-separating, and simultaneous degenerations.
The resulting tables summarize local monodromy types and
their reductive centralizers. We then discussed a particular
class of these labels, selected by imposing
\(\mathfrak h_i\subseteq\mathfrak c_i\) separately on each tube.

We conclude by discussing  several directions for further study:
\begin{itemize}

\item \textbf{Monodromy-sensitive physical observables.}
A natural question is which observables could probe the
tube monodromy data \((M_s,e)\). Candidates
include local-operator correlation functions, supersymmetric line
and defect observables, and subleading terms in \(Z_{S^4}\) and the
Zamolodchikov metric. The relation between line operators and
flat-connection data makes line observables particularly natural
candidates
\cite{Gaiotto:2010be,GaiottoMooreNeitzkeSpectralNetworks,Coman:2015lna}.

\item \textbf{AGT and sewing.}
The \(A_1\) four-punctured-sphere theory provides a controlled setting
for comparing localization with AGT/Liouville sewing
\cite{Alday:2009aq}. It would be interesting to determine whether
the tube monodromy data admit an interpretation in the corresponding
conformal blocks, correlation functions, or defect insertions
\cite{Teschner:2013tqy,Coman:2015lna}.

\item \textbf{Duality and extensions.}
It would be useful to determine how the cusp labels transform
under the mapping-class-group action relating weak-coupling frames,
and to investigate possible counterparts for Langlands-dual
Hitchin systems
\cite{Gaiotto:2009we,Kapustin:2006pk,Witten:2009at}.
Extensions to type \(D\) and \(E\) theories are also natural.

\item \textbf{Global protected-operator geometry.}
Another direction is the global geometry of protected operators
over conformal manifolds. Localization and the four-dimensional
\(tt^*\) equations provide local information, while Berry transport
and duality describe how these data are related globally
\cite{Papadodimas:2010tt,Baggio:2015tt,Baggio:2017berry}.
It would be interesting to investigate whether the local tube
monodromy data considered here are reflected in the asymptotic
geometry of these protected-operator bundles.

\item \textbf{Asymptotic non-Abelian Hodge theory.}
A mathematical direction is to formulate an asymptotic
non-Abelian Hodge description of degenerating families of harmonic
bundles and their associated local systems, and to relate their
limiting behavior to the tube monodromies used here
\cite{simpson1990harmonic,Mochizuki2003Asymptotic,mochizuki2009kobayashi}.

\item \textbf{Relation to the CFT Distance Conjecture.}
The infinite-distance behavior of weak-coupling cusps connects
this analysis to the CFT Distance Conjecture
\cite{Baume:2020dqd,Perlmutter:2020buo,Baume:2023msm,
Calderon-Infante:2024oed,CalderonInfante:2026quivers,MantegazzaEtAl2026,BaumeMantegazza2026,Fenati:2026hkn}.
A natural question is whether the additional monodromy decorations
refine the asymptotic characterization of these limits beyond the
leading Zamolodchikov metric.

\item \textbf{Global class-\(\mathcal S\) realizations.}
It would be useful to compare the local non-Abelian monodromy data with the
global Hitchin-system description near degenerations of the UV
curve and clarify the dictionary between them. In particular, one
may ask which local decorations arise from flat connections
associated with global Hitchin-system data compatible with the
puncture data and the relevant regularity and stability conditions,
and how these realizations relate to the fixture-and-gluing
description of the weak-coupling channel.

\end{itemize}

\section*{Acknowledgments}
We thank José Calderón-Infante and Irene Valenzuela for discussions. We thank Jacques Distler for comments that led us
to make the distinction between the particular class
considered in Section~6.3 and the general construction explicit. A.M.~thanks the organizers and participants of \emph{Pro(v/b)ing the Swampland}, where this work was presented, for their questions and comments. A.M.~thanks CERN-TH for hospitality and financial support during the final stages of this work. The work of A.M.~was supported in part by a grant from the Simons Foundation (602883, CV) and the DellaPietra Foundation. The work of T.G.~was carried out largely during a research stay at Harvard University, where he visited CMSA and the Swampland Initiative, whose hospitality and support he gratefully acknowledges. T.G.~is also supported by an NWO Vici grant.

\paragraph{Supplementary NAH audit companion.}
We supplement this paper with an AI-assisted non-Abelian Hodge (NAH) audit
companion developed for this work. Its checks were carried
out during manuscript finalization, after the
authors had developed the scientific results and arguments presented
here. Its role was confined to a separate cross-check of the mathematical details;
it played no role in developing or deriving those results.
The companion checks the hypotheses, conventions, and applicability
of the mathematical results underlying our NAH applications against
primary mathematical sources by Simpson, Corlette, Mochizuki, and
Deligne
\cite{simpson1990harmonic,Simpson1992HiggsBA,Corlette1988flat,
Mochizuki2002,Mochizuki2003Asymptotic,mochizuki2004kobayashi,
mochizuki2009kobayashi,deligne2006equations}.

The companion provides:
\begin{itemize}
    \item \textbf{Permanent, reusable checkpoints.}
    A knowledge base recording precise source references, mathematical
    statements, hypotheses, and limitations.

    \item \textbf{Claim tracking and audit history.}
    A map of manuscript claims and their relationships as stated in the text, constructed by auditing the manuscript, together with dated audit reports preserved across revisions.

    \item \textbf{Two-level reporting.}
    Guided by \emph{mathematical completeness and physics-facing
    reporting}, the companion preserves the full mathematical
    assessment and separately reports findings to the authors in
    language and at a level of detail appropriate to a
    theoretical-physics paper.

    \item \textbf{Reuse and extension.}
    The workflow can be reused for other NAH-related manuscripts by
    identifying the relevant primary mathematical sources and extending
    the permanent checkpoint database. It can also be adapted to other
    mathematical topics.
\end{itemize}

None of the scientific content of the paper was generated by AI. All decisions concerning the manuscript were made by the authors,
who retain responsibility for its content.
The earliest manuscript audit retained in our records is dated
4 August 2026. The companion is available at
\url{https://github.com/ammohseni/nah-audit-companion}.
 
\appendix

\section{Higgs bundles and associated vector bundles}
\label{app:higgs-bundles}

We briefly review the vector-bundle conventions underlying the
Hitchin-system and non-Abelian Hodge descriptions used in the main text.

\paragraph{Vector bundle.}
Let \(V\to C\) be a holomorphic rank-\(N\) vector bundle. At each point
\(p\in C\), its fiber \(V_p\) is an \(N\)-dimensional complex vector space,
locally identified with \(\mathbb C^N\). On a patch \(U_i\subset C\), the
bundle is trivial,
\be
    V|_{U_i}\simeq U_i\times\mathbb C^N.
\ee
After choosing a local basis
\(\{e_a^{(i)}\}_{a=1}^N\), a section takes the form
\be
    s(z)
    =
    \sum_{a=1}^N
    s_i^a(z)e_a^{(i)}(z).
\ee
On overlaps \(U_i\cap U_j\), the local descriptions are related by transition
functions \(g_{ij}\). For structure group \(SL(N,\mathbb C)\),
\be
    g_{ij}(z)\in SL(N,\mathbb C),
\ee
and the determinant of \(V\) is trivial.

\paragraph{Higgs field.}
A Higgs bundle is a pair \((V,\Phi)\), where \(\Phi\) is an
endomorphism-valued one-form. The endomorphism bundle \(\End(V)\) has fiber
\be
    \End(V)_p
    =
    \End(V_p)
    =
    \Hom(V_p,V_p),
\ee
while the canonical bundle \(K_C\) is the holomorphic cotangent bundle of
\(C\). Its local sections are holomorphic one-forms \(f(z)\,dz\). On a curve
without punctures, the Higgs field is therefore
\be
    \Phi
    \in
    H^0\!\left(
        C,
        \End(V)\otimes K_C
    \right),
\ee
where \(H^0\) denotes the space of globally defined holomorphic sections.
Equivalently, it is a bundle map
\be
    \Phi:
    V
    \longrightarrow
    V\otimes K_C.
\ee
Locally,
\be
    \Phi=\phi(z)\,dz,
    \qquad
    \phi(z)\in\End(V_z).
\ee
Thus \(\Phi\) acts on the internal fiber of \(V\) while also carrying a
holomorphic one-form on the curve.

For a punctured curve, let
\(
    D=\sum_a p_a
\)
denote the divisor of punctures. The twisted canonical bundle
\be
    K_C(D)=K_C\otimes\mathcal O_C(D)
\ee
allows one-forms with at most simple poles at the points \(p_a\). In the
\(SL(N,\mathbb C)\) theory, the Higgs field is traceless and hence satisfies
\be
    \Phi
    \in
    H^0\!\left(
        C,
        K_C(D)\otimes\End_0(V)
    \right),
    \qquad
    \Tr\Phi=0,
\ee
where \(\End_0(V)\) is the bundle of traceless endomorphisms of \(V\).
Accordingly, \(\Phi\) is a globally defined traceless matrix-valued
one-form, holomorphic away from the punctures and allowed to have simple
poles along \(D\).

\paragraph{From a principal bundle to a vector bundle.}
Let \(P\to C\) be a holomorphic principal \(G_{\mathbb C}\)-bundle. Given a
finite-dimensional representation
\be
    \rho:
    G_{\mathbb C}
    \longrightarrow
    \operatorname{GL}(W),
\ee
the group-valued transition functions of \(P\) act linearly on \(W\) and
define the associated vector bundle
\be
    V=P\times_{\rho}W.
\ee
More explicitly, this construction identifies
\be
    (p,w)
    \sim
    \bigl(pg,\rho(g)^{-1}w\bigr),
    \qquad
    g\in G_{\mathbb C}.
\ee
For the type-\(A\) theories considered here,
\(G_{\mathbb C}=SL(N,\mathbb C)\) and \(W=\mathbb C^N\) is the defining
representation, producing the rank-\(N\) vector bundle used above.

\section{Theta functions, half-differentials, and the prime form}
\label{app:theta-prime-form}

In this appendix, we collect the definitions of the functions used in the discussion of relative periods. The Riemann theta function is defined by
\be
    \theta(u\mid \Omega)
    =
    \sum_{n\in\mathbb Z^g}
    \exp\left[
        \pi i\, n^T\Omega n
        +
        2\pi i\, n^T u
    \right] .
\ee
A theta characteristic is a pair
\be
    \Delta=(\Delta',\Delta''),
    \qquad
    \Delta',\Delta''\in \frac12\mathbb Z^g/\mathbb Z^g .
\ee
Here \(\Delta'\) and \(\Delta''\) are half-integer vectors defined modulo integer shifts; equivalently, each component is either \(0\) or \(\frac12\) modulo \(\mathbb Z\). The theta function with characteristic \(\Delta\) is
\be
    \theta[\Delta](u\mid\Omega)
    =
    \sum_{n\in\mathbb Z^g}
    \exp\left[
        \pi i (n+\Delta')^T\Omega(n+\Delta')
        +
        2\pi i (n+\Delta')^T(u+\Delta'')
    \right].
\ee
The characteristic is called even or odd according to the parity (even or odd) of
\(
    4\,\Delta'^T\Delta'' \, {\rm mod}\; 2.
\)
Equivalently,
\be
    \theta[\Delta](-u\mid\Omega)
    =
    \pm \theta[\Delta](u\mid\Omega),
\ee
with the plus sign for an even characteristic and the minus sign for an odd
characteristic.

To define the prime form, choose a nonsingular odd theta characteristic
\(\Delta\). Since \(\Delta\) is odd, one has
\(
    \theta[\Delta](0\mid\Omega)=0
\).
Nonsingular means that the gradient at the origin is nonzero,
\(
    \nabla_u\theta[\Delta](0\mid\Omega)\neq 0 
\).
The corresponding half-differential \(h_\Delta(p)\) is defined locally by
\be
    h_\Delta(p)^2
    =
    \sum_{i=1}^g
    \frac{\partial \theta[\Delta](0\mid\Omega)}{\partial u_i}\,
    \omega_i(p).
\ee
Here \(\omega_i(p)\) denotes the value of the holomorphic one-form \(\omega_i\) at the point \(p\). The right-hand side is a holomorphic one-form in \(p\). Hence \(h_\Delta(p)\) is
locally a square root of a holomorphic one-form.

The prime form is
\be
    E(p,q\mid\Omega)
    =
    \frac{\theta[\Delta](u_{pq}\mid\Omega)}
    {h_\Delta(p)h_\Delta(q)} .
\ee
Although this formula uses the auxiliary odd characteristic \(\Delta\), the
resulting prime form is independent of this choice up to the standard overall
sign convention. It is antisymmetric,
\be
    E(p,q\mid\Omega)=-E(q,p\mid\Omega),
\ee
and has a simple zero along the diagonal \(p=q\), with no other zeros.

More explicitly, if \(z\) is a local coordinate near \(p=q\), then
\be
    E(p,q\mid\Omega)
    =
    \frac{z(p)-z(q)}
    {\sqrt{dz(p)}\sqrt{dz(q)}}
    \left(
        1+O\bigl((z(p)-z(q))^2\bigr)
    \right).
\ee
Thus, after choosing local coordinates and trivializing the half-differentials,
one often writes simply
\be
    E(p,q\mid\Omega)\sim z(p)-z(q).
\ee
This is the sense in which the prime form gives a local separating coordinate for
the collision of two marked points.

\subsection{Elliptic specialization and modular transformations}\label{thetafunction}

Let
\(
    E_{\tau_E}=\mathbb C/(\mathbb Z+\tau_E\mathbb Z)
\)
be an elliptic curve. We use the odd Jacobi theta function
\be
    \vartheta_1(z\mid \tau_E)
    =
    -i\sum_{n\in\mathbb Z}
    (-1)^n
    \exp\!\left[\pi i\left(n+\frac12\right)^2\tau_E\right]
    \exp\!\left[2\pi i\left(n+\frac12\right)z\right].
\ee
It is an odd holomorphic function of \(z\), with a simple zero at \(z=0\)
\be
    \vartheta_1(-z\mid\tau_E)=-\vartheta_1(z\mid\tau_E),
    \qquad
    \vartheta_1(z\mid\tau_E)
    =
    \vartheta_1'(0\mid\tau_E)\,z+O(z^3).
\ee
Under shifts by the elliptic lattice, it transforms as
\ba
    &\vartheta_1(z+1\mid\tau_E)
    =
    -\vartheta_1(z\mid\tau_E),\\
   & \vartheta_1(z+\tau_E\mid\tau_E)
    =
    -e^{-\pi i\tau_E-2\pi i z}
    \vartheta_1(z\mid\tau_E).
\ea
Thus \(\vartheta_1(z\mid\tau_E)\) is not an ordinary function on the quotient
\(E_{\tau_E}\); rather, it is covariant under the lattice identifications.

The prime form on the elliptic curve can be written, in this normalization, as
\be
    E(z_1,z_2\mid\tau_E)
    =
    \frac{\vartheta_1(z_1-z_2\mid\tau_E)}
    {\vartheta_1'(0\mid\tau_E)} .
\ee
Near the diagonal \(z_1=z_2\), one has
\(
    E(z_1,z_2\mid\tau_E)
    =
    z_1-z_2+O\!\left((z_1-z_2)^3\right).
\)
Thus, in the notation of Section~\ref{sec:nah-asymptotics}, the separating
coordinate is locally
\be
    q_s \sim E(z_1,z_2\mid\tau_E)
    \sim z_1-z_2 .
\ee

Under modular transformations of the complex structure,
\[
    \tau_E\longmapsto \frac{a\tau_E+b}{c\tau_E+d},
    \qquad
    \begin{pmatrix}
    a & b \\
    c & d
    \end{pmatrix}
    \in SL(2,\mathbb Z),
\]
the theta function transforms covariantly. In particular, for the generators
\(T:\tau_E\mapsto \tau_E+1\) and \(S:\tau_E\mapsto -1/\tau_E\), one has
\ba
   & \vartheta_1(z\mid\tau_E+1)
    =
    e^{\pi i/4}\vartheta_1(z\mid\tau_E),\\
   & \vartheta_1\!\left(\frac{z}{\tau_E}\,\middle|\,-\frac{1}{\tau_E}\right)
    =
    -i(-i\tau_E)^{1/2}
    e^{\pi i z^2/\tau_E}
    \vartheta_1(z\mid\tau_E).
\ea
Thus \(\vartheta_1\) is not a modular form in the usual sense; it is a
Jacobi theta function, transforming with both a multiplier and an exponential
factor.

\bibliographystyle{jhep}
\bibliography{references}
\end{document}